\documentclass[journal]{IEEEtran}

\usepackage{bm}
\usepackage[cmex10]{amsmath}
\usepackage{amssymb}
\usepackage[pdftex]{graphicx}
\usepackage{stfloats}
\usepackage{algorithm}
\usepackage{algorithmic}
\usepackage[numbers,sort&compress]{natbib}

\usepackage{color}
\newenvironment{proof}{{\indent\it Proof: \ }}{\hfill $\blacksquare$\par}

\makeatletter
\newcommand{\biggg}{\bBigg@{3}}

\newcommand{\Biggg}{\bBigg@{3.5}}

\newcommand{\bigggg}{\bBigg@{4}}

\newcommand{\Bigggg}{\bBigg@{4.5}}

\makeatother

\ifCLASSINFOpdf

\else

\fi

\begin{document}

\title{Adaptive Channel Estimation and Hybrid Beamforming for RIS aided Vehicular Communication}

%
%
%
\author{Tianyou Li, Haifeng Hu, Dapeng Li,~\IEEEmembership{Member,~IEEE,}

  
}

%
%

\markboth{Journal of \LaTeX}%
{Shell \MakeLowercase{\textit{et al.}}: Bare Demo of IEEEtran.cls for IEEE Journals}
%



\maketitle

\begin{abstract}
\small {Reconfigurable intelligent surface (RIS) constitutes a disruptive technology for enhancing vehicular communication performance through reconfigurable propagation environments. In this paper, we propose an adaptive channel estimation framework and hybrid beamforming optimization strategy for RIS-aided vehicular multiple-input multiple-output (MIMO) systems operating in high-mobility scenarios. To address severe Doppler effects and rapid channel variations, we design a velocity-aware pilot scheme that progressively estimates cascaded channels across two timescales, leveraging tensor decomposition and adaptive grouping of passive elements. This framework dynamically balances channel estimation accuracy and spectral efficiency, significantly reducing training overhead. Furthermore, we develop a low-complexity hybrid beamforming algorithm for both narrowband single vehicle user equipment (VUE) and broadband multi-VUE systems. For single-VUE scenarios, we derive closed-form active beamforming solutions and optimize passive beamforming via alternating optimization. For multi-VUE broadband systems, we jointly optimize subcarrier allocation, power distribution, and beamforming to maximize system throughput while mitigating inter-carrier interference (ICI) caused by Doppler spread, subject to quality-of-service (QoS) constraints and RIS hardware limitations. Our simulation results demonstrate that the proposed methods achieve substantial performance gains in channel estimation efficiency, beamforming robustness, and system throughput compared to conventional schemes, particularly under high mobility conditions.}
\end{abstract}


%
\IEEEpeerreviewmaketitle

\vspace{-4mm}
\section{Introduction}

\IEEEPARstart{I}{NTELLIGENT} transportation is a modern transportation management system supported by next-generation communication technology, artificial intelligence, the Internet of Things, and big data technology, characterized by information connectivity, real-time monitoring, and collaborative management \cite{art1, art2}. 
Intelligent transportation is a key solution to addressing the imbalance between supply and demand, efficiency bottlenecks, and ecological constraints in traditional transportation systems. 
By 2030, intelligent transportation could reduce urban traffic congestion by $30\%$ and lower traffic accident rates by over $20\%$, making its scientific value and application potential a global competitive focus \cite{2024Intelligent1}. 
The core of achieving smart transportation lies in utilizing next-generation communication systems as the neural hub of smart transportation.
By leveraging their capabilities for high-capacity, high-reliability, low-latency, and wide-coverage information transmission, these systems support integrated control across vehicles, roads, and clouds, driving a paradigm shift in transportation vehicles from \emph{single-vehicle intelligence} to \emph{multi-vehicle collaborative intelligence}.

To achieve high-capacity, high-reliability, low-latency, and wide-coverage information transmission capabilities, reconfigurable intelligence surface (RIS) technology, as the next-generation mobile communication technology, has garnered considerable attention \cite{art4, wei2025machine}.
Deploying a large number of passive or active elements on a RIS enables the adjustment of the phase and amplitude of incident signals by controlling the elements, thereby improving the signal transmission environment and enhancing signal quality. 
Transparent RIS is a novel intelligent metasurface technology that allows signals to pass through without being reflected to the other side, making it easy to deploy on the outer surface of objects and enhancing the communication performance of the receiver on the same side as the RIS \cite{10546994, 10737393}. 
Currently, the continuous growth of autonomous driving chip computing power has made real-time control of in-vehicle transparent RIS possible, demonstrating significant potential in the field of intelligent transportation communications \cite{odema2024performance, narashiman2024chiplets}. 

However, the introduction of in-vehicle transparent RIS technology has also brought a series of new challenges to intelligent transportation communication systems.
First, RIS has led to a sharp increase in the number of channel state information (CSI) dimensions, which has also caused the problem of excessive channel estimation overhead to become increasingly prominent, especially in high-speed scenarios. 
Furthermore, in transparent RIS-aided communication systems, traditional channel estimation algorithms cannot be applied due to the passive reflection characteristics of RIS. 
\cite{shao2022target} constructed a new self-sensing framework based on RIS, equipping RIS with dedicated low-cost sensors to receive signals, thereby promoting channel estimation and target localization in communications.
\cite{wang2023joint} proposed a RIS-enhanced joint positioning and communication framework that deployed sensors on RIS to obtain CSI to optimize the RIS phase shift matrix, thereby improving positioning accuracy and communication speed.
However, the above research introduces new complexity to RIS by deploying sensors, which runs counter to the vision of RIS as passive and low-cost.
\cite{chen2023channel} and \cite{hu2021two} utilized the characteristic that the sparse channel matrix of all user cascade channels has a common block sparse structure, and proposed a two-stage channel estimation framework that effectively reduces the pilot overhead.
\cite{shen2023deep,rahman2023deep,9611281,10556747,11020694} used deep learning algorithms to estimate cascaded channels and achieved better performance than traditional channel estimation algorithms.
However, the research scenarios  of the existing channel estimation methods in RIS-aided communication systems are mostly stationary or low-speed, and the mobility issues of in-vehicle RIS are not considered. 
High-speed moving transmitter and receiver not only cause severe Doppler effects, but also lead to a decrease in channel estimation efficiency.

Second, beamforming algorithms in RIS-aided vehicular communication systems also lack in-depth research and application.
Beamforming in RIS-aided communication systems includes not only active beamforming on the transmitter side, but also reflective passive beamforming on the RIS side, which is called hybrid beamforming.
Due to the large number of optimization parameters in hybrid beamforming and their mutual coupling, it is difficult to optimize them individually. 
The current research approach is to decouple the hybrid beamforming optimization problem into two separate subproblems, and then use non-convex optimization algorithms to perform alternating iterative optimization on the two subproblems, continuously approaching a local optimal solution, thereby improving the communication performance of the system \cite{9110912,lyu2021hybrid,li2021sum}.
\cite{10994825,10879255} employed manifold optimization techniques to jointly design active beamforming at the base station (BS) and passive beamforming at the RIS under manifold constraints, with the aim of maximizing transmission rates.
\cite{xiang2025deep,tang2025joint,bhardwaj2025fault} addressed the optimization problem of hybrid beamforming by utilizing deep learning framework and combining methods such as offline training and online prediction to optimize beam alignment time and communication performance.
However, it is worth noting that existing RIS-aided communication system beamforming algorithms generally assume perfect CSI and do not consider the difficulty of obtaining CSI in intelligent transportation scenarios, making the proposed beamforming algorithms difficult to apply in practice.
In addition, existing methods have not taken into account the impact of Doppler interference on the system in transportation communications.
More importantly, there has been a lack of in-depth research on resource allocation issues in vehicular communication systems with multiple vehicle user equipments (VUEs).

In this paper, we study the channel estimation method with adaptive pilot and hybrid
beamforming optimization in RIS aided vehicular communication MIMO system for maximizing the performance of single-VUE and multi-VUE. 
Our main contributions are summarized as follows.
\begin{itemize}
\item First, we design an adaptive channel estimation framework for RIS-aided high-mobility vehicular MIMO systems. 
Parallel factorization is used for three-dimensional tensor information processing in MIMO systems.
By leveraging the two-timescale channel properties, a novel estimation scheme is developed that adaptively balances the trade-off between channel estimation accuracy and transmission efficiency through velocity-aware pilot design, and then progressively estimates channel state information across both timescales.
The proposed method significantly reduces training overhead and accommodates rapid channel variations.
\item Next, we develop a grouped hybrid beamforming method tailored to aggregated CSI blocks for single-VUE in narrowband system, significantly reducing complexity via phase group optimization while maintaining near-optimal rate performance. 
We also derive the optimal closed-form solution and descending gradient for different cases.
The proposed algorithm can achieve progressive beamforming optimization based on imperfect group estimation channels, thereby improving system transmission rates and reducing communication transmission latency.
\item Moreover, we study the resource allocation problem for multi-VUE in broadband system.
Unlike most works that neglect inter-carrier interference (ICI) induced by high mobility, we explicitly model ICI in subcarrier allocation and jointly optimize power, beamforming with QoS, RIS hardware constraints.
The problem is decoupled into the resource allocation problem and the beamforming problem.
Based on the auxiliary integer relaxation and approximation from difference of convex (D.C.), the resource allocation problem is transformed into convex one.
It can be solved iteratively with the beamforming problem by using convex optimization algorithms.
\item Finally, we provide extensive numerical results to validate the performance advantages of our proposed adaptive channel estimation framework, beamforming optimization and resource allocation algorithms. 
In particular, it is shown that the proposed channel estimation framework can flexibly change the number of pilot without losing too much performance. 
Compared with other conventional schemes, the proposed beamforming optimization and resource allocation algorithms have more performance improvements.

\end{itemize}

The remainder of this paper is organized as follows: Section II introduces the system model of transparent RIS aided vehicular communication system and information transmitting model. 
In Section III, we propose the adaptive channel estimation framework in two timescales for progressively channel estimation. 
In  Section IV, based on the estimated cascaded CSI, we optimize the beamforming for performance of vehicular communication system with single-VUE and multi-VUE respectively. 
Then, numerical simulation results and their pertinent discussions are presented in Section V. 
Finally, the paper is ended with conclusion in Section VI.


\vspace{-4mm}
\section{System Model}
\vspace{-2mm}
\begin{figure}{}
\centering
  \includegraphics[width=3.5in,height=1.9in]{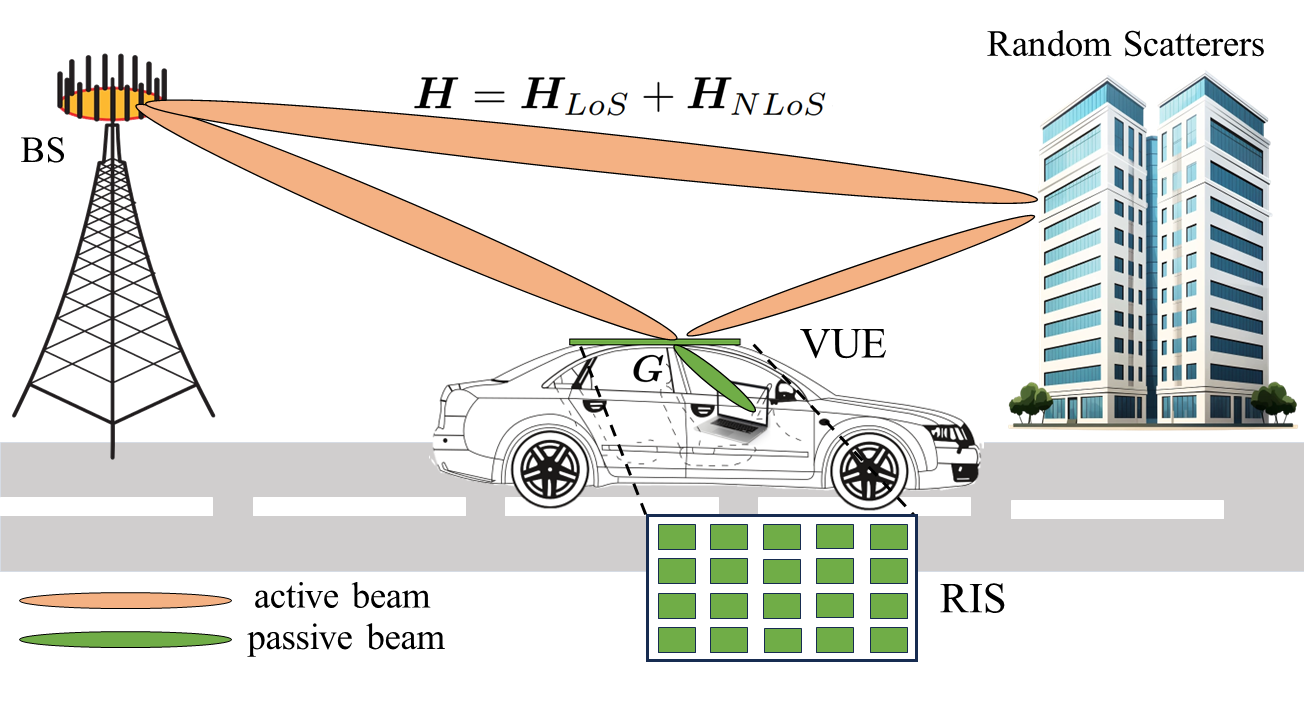}
\vspace{-8mm}
\caption{A transparent RIS is deployed on the roof of the vehicle and multiple antenna are deployed at both sides of transmitter and receiver.
The RIS can be controlled by the in-vehicle controller via the wireless
 backhaul link.
}
\vspace{-6mm}\label{figure1}
\end{figure}
In this section, we elaborate on the system model from both channel estimation phase and information transmission phase in the high-mobility vehicular MIMO communication system aided by one RIS in the urban areas.
As depicted in Fig. 1, the BS with $N_t \geq 1$ antennas is deployed on the roadside
The vehicle runs on the straight road at the speed of $v$ meters/second (m/s).
Considering that the interior of the vehicle is large enough, the VUE is assumed to be equipped with $N_r \geq 1$ antennas.
One transparent RIS equipped with $M \geq 1$ passive reflecting elements is deployed at the top of the vehicle to replace the metal panel that incurs high penetration loss and aid VUE's communication \cite{LOSNLOS}.
The set of passive elements in RIS is defined as $\mathcal{M} = \{1, 2,\dots, M\}$.
The signal transmitted by the roof side BS can penetrate the transparent RIS and be dynamically adjusted by the RIS controller via separate reliable wireless backhaul links for desired signal reflection.
Specifically, the RIS can add a phase shift to the signal incident on each passive element and adjust the amplitude.
Let $\phi_m \in (0, 2\pi]$ $\forall m=1,\dots,M$ denotes the phase shift introduced to the incident signal passing the $m$-th passive element by the RIS.
The reflection coefficient of the $m$-th element is expressed as $\theta_m=e^{j\phi_m}$.
As for the amplitude reflection coefficient of the RIS, we simply set it to 1 for all reflection elements, which has been proved to be optimal in \cite{amplitude}.
In this work, we focus on the physical-layer design of adaptive channel estimation and hybrid beamforming for RIS-aided vehicular communication systems.
To maintain analytical tractability and highlight the proposed algorithmic contributions, we assume that the vehicle remains connected to a single serving BS within the considered coverage area, and the handover process across cells is not considered.

Next, the channel model will be introduced.
The downlink baseband channels from the BS to the RIS, the RIS to the VUE are denoted by $\bm{H} \in \mathbb{C}^{M\times N_t}$, $\bm{G} \in \mathbb{C}^{N_r\times M}$, respectively. 
Considering that the BS is deployed at the building top, we assume that the BS-RIS channel $\bm{H}$ is LoS-dominant, which is modeled as the following Rician fading channel 
\begin{align}
\bm{H}=\bm{H}_{LoS}+\bm{H}_{NLoS}.
\end{align}
We assume that the LoS and the NLoS components follow the distribution of CSCG, i.e., $\text{vec}(\bm{H}_{LoS})\sim \mathcal{CN}(\bm{0},\frac{P_{L}\mathcal{K}}{1+\mathcal{K}}\bm{I}_{N_tM})$, $\text{vec}(\bm{H}_{NLoS})\sim \mathcal{CN}(\bm{0},\frac{P_{L}}{1+\mathcal{K}}\bm{I}_{N_tM})$, where $P_{L}$ is the large-scale fading factor and $\mathcal{K}$ is the Rician factor.
Similarly, the RIS-VUE channel $\bm{G}$ is also modeled as the Rician fading channel, i.e., $\text{vec}(\bm{G})\sim \mathcal{CN}(\bm{0},P_{L}\bm{I}_{MN_r})$.
It is worth noting that due to the high penetration loss incurred by the side metal panel of the vehicle, the amplitude of the direct signal from the BS to the VUE is small and ignored.
To improve the performance of the system, we can pre-estimate the direct link by turning off the RIS, which is well mature in academia. In order to simplify the exposition of the channel estimation of the RIS, the signal transmitted by the direct link can be neglected.
Moreover, the BS-RIS LoS channel $\bm{H}_{LoS}$ can be considered as approximately constant within several transmission frames of interest.
This is practically reasonable since the traveling distance of the VUE within several transmission frames can be neglected as compared to the nominal distance with the BS \cite{9661068}.
As a result, there are only small variations in the geometry-related channel parameters, including the Doppler frequency.
As for the RIS-VUE channel $\bm{G}$, due to the short distance between the transparent RIS and the VUE as well as the fact that they remain relatively static, it varies more slowly as compared to $\bm{H}_{LoS}$.
Consequently, $\bm{H}_{LoS}$ and $\bm{G}$ can be viewed as invariant at the period of interest \cite{static1}.

In the channel training phase, the BS transmits pilot signals to the VUE aided by the transparent RIS.
The received signal vector $\bm{y}\in\mathbb{C}^{N_r\times1}$can be presented as follows
\begin{align}
\bm{y}=\bm{G\Phi Hx}+\bm{n},
\end{align}
where $\bm{x}\in\mathbb{C}^{N_t\times1}$ is the channel estimation pilot signal vector, $\bm{\Phi}=\text{diag}\left\{\theta_1,\dots,\theta_M\right\}$.
Note that $\bm{n}\in\mathbb{C}^{Nr\times1}$ is the additive white Gaussian noise (AWGN) vector, where $\{n_i\}_{i=1}^{N_r}\sim\mathcal{CN}(0,\sigma^2)$. 
With the estimated CSI, the BS transmits information to the VUE in the downlink information transmission phase. 

As mentioned above, the transparent RIS lacks information processing capabilities due to the passive reflection, thus channel estimation for the proposed vehicular communication system is challenging.
Existing channel estimation algorithms cannot be applied to the considered system directly.
\vspace{-3mm}
\section{Adaptive Channel Estimation Framework}
In this section, the adaptive channel estimation is introduced. 
Firstly, we propose a typical generic estimation protocol.
Then, the channel training process is divided into two phases from a timescale perspective.
Last, with our proposed channel estimation algorithm, we progressively obtain CSI on two timescales.
The details of the proposed channel estimation framework of the communication system aided by transparent RIS are given in the following.

As discussed in Section II, the coherence time of $\bm{H}_{LoS}$ and $\bm{G}$ is much longer than that of $\bm{H}_{NLoS}$, which can be presented as $T^{coh}_{LoS}/T^{coh}_{N\!LoS}\gg 1$. 
$T^{coh}_{LoS}$ is the coherence time of $\bm{H}_{LoS}$ and $\bm{G}$, while $T^{coh}_{N\!LoS}$ is the coherence time of $\bm{H}_{NLoS}$.
The small timescale channel coherence time $T^{coh}_{N\!LoS}$ is divided into estimation phase and transmission phase.
The estimation phase is further divided into $I$ blocks, where each block has $T$ time slots.
Accordingly, there are $IT$ slots in total in one small timescale coherence time.
In one block, the pilot signals $\bm{x}_{i,1},\bm{x}_{i,2},\dots,\bm{x}_{i,T}$ are different, while they are constant and repeated over $I$ blocks.
As for the pilot phase shifts $\bm{\theta}_{1,t},\bm{\theta}_{2,t},\dots,\bm{\theta}_{I,t}$, they are different over blocks and constant in all time slots during one block.
\vspace{-5mm}
\subsection{Small Timescale Channel Estimation}
In this subsection, we introduce the framework of our proposed adaptive small timescale channel estimation.
Under the proposed protocol, the received signal model in $t$-th time slot of $i$-th block can be written as 
\begin{align}
\bm{y}_{i,t}=\bm{G\Phi}_{i}\bm{Hx}_{t}+\bm{n}_{i,t},\ \ &t=1,2,\dots,T,\notag\\ &i=1,2,\dots,I.
\end{align}
By stacking all $\bm{y}_{i,t}$ as the column vectors of $\bm{Y}_i\in\mathbb{C}^{N_r\times T}$, we can get the received signal matrix in $i$-th block as follows
\vspace{-3mm}
\begin{align}
\bm{Y}_i=\bm{G\Phi}_{i}\bm{HX}+\bm{N}_i,\ \ i=1,2,\dots,I,
\end{align}
where $\bm{X}=\left[\bm{x}_1,\dots,\bm{x}_T\right]\in\mathbb{C}^{N_t\times T}$, $\bm{N}_i=\left[\bm{n}_{i,1},\dots,\bm{n}_{i,T}\right]\in\mathbb{C}^{N_r\times T}$.
Similarly, by stacking all matrices $\bm{Y}_i$, we can get a three-way tensor $\bm{\mathcal{Y}}\in\mathbb{C}^{N_r\times T\times I}$.
The matrix $\bm{Y}_i$ is $i$-th frontal matrix slice of $\bm{\mathcal{Y}}$, whose $(n_r,t,i)$-th entry can be presented as $\bm{\mathcal{Y}}_{n_r,t,i}=\sum^{M}_{m=1}g_{n_r,m}\theta_{i,m}\bm{h}_{m}\bm{x}_t$, where $g_{n_r,m}$ is the $n_r$-th row, $m$-th column entry of $\bm{G}$ and $\bm{h}_{m}$ is the $m$-th row of $\bm{H}$.
In order to simplify the process of the received pilot signals, by exploiting the trilinearity of the parallel factor (PARAFAC) decomposition, the received tensor $\bm{\mathcal{Y}}$ can be unfolded in the form of matrix as follows \cite{9104260, 10974631}:
\begin{align}
\bm{\tilde{Y}}=\bm{\Xi}^T\left(\left(\bm{HX}\right)^T\odot\bm{G}\right)^T+\bm{\tilde{N}}\in\mathbb{C}^{I\times N_rT}\label{pilotY},
\end{align}
where $\bm{\tilde{Y}}\triangleq\left[\text{vec}\left(\bm{Y}_1\right),\dots,\text{vec}\left(\bm{Y}_I\right)\right]^T$, 
$\bm{\Xi}=\left[\bm{\theta}_1,\dots,\bm{\theta}_I\right]$,
$\bm{\tilde{N}}\triangleq\left[\text{vec}\left(\bm{N}_1\right),\dots,\text{vec}\left(\bm{N}_I\right)\right]^T$.
By applying the property $(\bm{AB})\odot(\bm{CD})=(\bm{A}\otimes\bm{C})(\bm{B}\odot\bm{D})$, we can recast (\ref{pilotY}) as
\begin{align}
\bm{\tilde{Y}}&=\bm{\Xi}^T\left(\left(\bm{X}^T\bm{H}^T\right)\odot\left(\bm{I}_{N_r}\bm{G}\right)\right)^T+\bm{\tilde{N}}\notag\\
&=\bm{\Xi}^T\left(\bm{H}^T\odot\bm{G}\right)^T\left(\bm{X}\otimes\bm{I}_{N_r}\right)+\bm{\tilde{N}}\label{pilotY2}.
\end{align}
By right-producting the inverse of the pilot signal matrix $\left(\bm{X}\otimes\bm{I}_{N_r}\right)^{-1}=\left(\bm{X}^{-1}\otimes\bm{I}_{N_r}\right)$ at both sides, the received unfolded signal can be rewritten as 
\begin{align}
\bm{\bar{Y}}\triangleq\bm{\tilde{Y}}\left(\bm{X}^{-1}\otimes\bm{I}_{N_r}\right)=\bm{\Xi}^T\left(\bm{H}^T\odot\bm{G}\right)^T+\bm{\bar{N}},\label{unfold}
\end{align}
where $\bm{\bar{N}}\triangleq\bm{\tilde{N}}\left(\bm{X}^{-1}\otimes\bm{I}_{N_r}\right)$.
Note that for the existence of $\bm{X}^{-1}$, $T\geq N_t$ is required.
With the received pilot signal $\bm{\bar{Y}}$, we can easily obtain the CSI of the cascaded channel $\bm{\bar{H}}\triangleq\left(\bm{H}^T\odot\bm{G}\right)^T\in\mathbb{C}^{M\times N_tN_r}$ by left product the inverse of the pilot phase matrix $\bm{\Xi}^{-T}$.
Same as $\bm{X}^{-1}$, to ensure the existence of $\bm{\Xi}^{-T}$, $I\geq M$ is needed.
As discussed above, at least $MN_r$ pilot symbols are required to get the accurate cascaded CSI.

However, the number of passive elements $M$ is usually very large in practice.
Moreover, due to the extremely short coherence time in the proposed high-mobility vehicular communication system, a large number of pilot symbols will lead to a significant reduction of the remaining information transmission time within the coherent block.
It may overwhelm the gain introduced by the RIS or even cause the communication system to collapse.
To address the issue, we propose an adaptive channel estimation method.
The method can adapt the number of pilot phase symbols that varies with the vehicle's velocity.
As the number of pilot phase symbols $I$ increases, the system can obtain progressively refined CSI.
By thoroughly designing the number $I$, the trade-off between the minimization of the channel estimation error and the maximization of the information transmission time can be obtained for achieving better system performance.

Next, we present the detailed method of the adaptive channel estimation. 
We define the group pilot phase in the $i$-th block as $\bm{\psi}^i\triangleq[\psi^i_1,\dots,\psi^i_i]$.
In the first block, the pilot phase vector $\bm{\theta}_1$ can be given as 
\begin{align}
\bm{\theta}_1=\left(\psi^1_1\otimes \bm{1}_{M\times 1}\right)^T.
\end{align}
The received signal in the first block can be represented as 
\begin{align}
\bm{\tilde{y}}_1=\left(\psi^1_1\otimes \bm{1}_{M\times 1}\right)^T\bm{\bar{H}}+\bm{\tilde{n}}_1.\label{ty1}
\end{align}
In the second block, by dividing the whole passive elements into two groups $\mathcal{M}^2_1$ and $\mathcal{M}^2_2$, the pilot phase can be decomposed into the Kronecker product of group pilot phase and the $\bm{1}$-vector. 
The received signal can be represented as
\begin{align}
\bm{\tilde{y}}_2&=
\begin{bmatrix}
\psi^2_1\otimes \bm{1}_{\left|\mathcal{M}^2_1\right|\times 1}\\
\psi^2_2\otimes \bm{1}_{\left|\mathcal{M}^2_2\right|\times 1}
\end{bmatrix}^{T}
\begin{bmatrix}
\bm{\bar{H}}_{1:\left|\mathcal{M}^2_1\right|}\\
\bm{\bar{H}}_{\left(\left|\mathcal{M}^2_1\right|+1\right):\left(\left|\mathcal{M}^2_1\right|+\left|\mathcal{M}^2_2\right|\right)}
\end{bmatrix}
+\bm{\tilde{n}}_2\notag\\
&=\begin{bmatrix}
\psi^2_1,\psi^2_2
\end{bmatrix}
\begin{bmatrix}
\bm{\bar{h}}^2_{agg,1}\\
\bm{\bar{h}}^2_{agg,2}
\end{bmatrix}+\bm{\tilde{n}}_2,\label{block2}
\end{align}
where $\bm{\bar{H}}_{a:b}$ denotes the rows of channel matrix from $a$-th to $b$-th.
The aggregated channel of the $k$-th $(k\leq i)$ group in $i$-th block is defined as
\begin{align}
\bm{\bar{h}}^i_{agg,k}&\triangleq\bm{1}_{1\times\left|\mathcal{M}^i_k\right| }\cdot\bm{\bar{H}}_{\left(\sum_{l=1}^{k-1}\left|\mathcal{M}^i_{l}\right|+1\right):\left(\sum_{l=1}^{k}\left|\mathcal{M}^i_{l}\right|\right)}.
\end{align}
By unwrapping the pilot phase $\psi^1_1$ in the first block, we have 
\begin{align}
\bm{\tilde{y}}_1=
\begin{bmatrix}
\psi^1_1,\psi^1_1
\end{bmatrix}
\begin{bmatrix}
\bm{\bar{h}}^2_{agg,1}\\
\bm{\bar{h}}^2_{agg,2}
\end{bmatrix}+\bm{\tilde{n}}_1.\label{block1}
\end{align}
Then, we can obtain the stacking received signal $\bm{\tilde{Y}}^2$ by combining (\ref{block2}) and (\ref{block1}) in the following
\begin{align}
\bm{\tilde{Y}}^2=
{\begin{bmatrix}
\psi^1_1,\psi^1_1\\
\psi^2_1,\psi^2_2
\end{bmatrix}}
{\begin{bmatrix}
\bm{\bar{h}}^2_{agg,1}\\
\bm{\bar{h}}^2_{agg,2}
\end{bmatrix}}+
\begin{bmatrix}
\bm{\tilde{n}}_1\\
\bm{\tilde{n}}_2
\end{bmatrix}.
\end{align}
In the $i$-th block, we can stack $i$ received signals and combine pilot phases in $(i-1)$ blocks and the $i$-th block by further dividing one of the element groups (generally the group with the largest number of elements) in $(i-1)$-th block. 
Accordingly, by dividing the $j$-th group $\mathcal{M}^{i-1}_{j}$ in $(i-1)$-th block $(j\leq i-1)$ into $\left\{\mathcal{M}^{i}_{j},\mathcal{M}^{i}_{j+1}\right\}$,
the group of elements can be given as $\mathcal{M}=\left\{ \mathcal{M}^{i-1}_1,\dots,\mathcal{M}^{i-1}_{j-1},\mathcal{M}^{i}_{j},\mathcal{M}^{i}_{j+1},\mathcal{M}^{i-1}_{j+1},\dots,\mathcal{M}^{i-1}_{i-1}\right\}$.
Therefore, we can obtain
\begin{align}
\mathcal{M}^{i}_{k}=
    \begin{cases}
    \mathcal{M}^{i-1}_{k},  \ \ &k\leq j-1,\\
    \mathcal{M}^{i-1}_{k-1}, &k\geq j+2.
    \end{cases}
\end{align}
By defining $\bm{\tilde{N}}^i\triangleq[\bm{\tilde{n}}_1,\dots,\bm{\tilde{n}}_i]^T,\bm{\bar{H}}_{agg}^i\triangleq[(\bm{\bar{h}}_{agg,1}^i)^T,\dots,(\bm{\bar{h}}_{agg,i}^i)^T]^T$ and constructing
\begin{align}
\bm{{\Psi}}^i=
{\begin{bmatrix}
\bm{{\Psi}}^{i-1}_{(1:j)},\bm{{\Psi}}^{i-1}_{j},\bm{{\Psi}}^{i-1}_{(j:(i-1))}\\
\bm{\psi}^i
\end{bmatrix}}\in\mathbb{C}^{i\times i},
\end{align}
the stacked received signal $\bm{\tilde{Y}}^i$ can be written as 
$
\bm{\tilde{Y}}^i=\bm{{\Psi}}^i\bm{\bar{H}}_{agg}^i+\bm{\tilde{N}}^i.
$
By properly designing the group pilot phase matrix $\bm{{\Psi}}^i$, the existence of inverse matrix $\left(\bm{{\Psi}}^i\right)^{-1}$ can be ensured for arbitrary $I\geq 1$ without requiring $I\geq M$.
By left-producting the inverse of the group pilot phase matrix, the estimated aggregated channel consisting of $I$ groups can be given as
\begin{align}
\hat{\bm{\bar{H}}}_{agg}^I=\left(\bm{{\Psi}}^I\right)^{-1}\bm{\tilde{Y}}^I=\bm{\bar{H}}_{agg}^I+\left(\bm{{\Psi}}^I\right)^{-1}\bm{\tilde{N}}^I.
\end{align}
The individual aggregated channel passing through the $i$-th group of passive elements can be obtained by de-vector the estimated channel as the $N_r\times N_t$ matrix.
As the number of pilot symbols increases, the resolution of the communication system to the channel increases and the error in channel estimation becomes reduced.
It is worth noting that when $I=M$, the $\hat{\bm{\bar{H}}}_{agg}^M$ is the complete estimated channel but not aggregated channel $\hat{\bm{\bar{H}}}$ since the number of elements in each group is 1.
Although it is impossible to obtain the channel of each individual element when $I<M$, we can still optimize the group phases of passive elements according to the estimated aggregated channel $\hat{\bm{\bar{H}}}_{agg}^I$, which will be introduced in the next section.
Using the proposed adaptive channel estimation method, we can design the number of training blocks in the coherence time based on the speed of the vehicle to achieve the trade-off between the minimization of the channel estimation error and the maximization of the information transmission time.
It means by sacrificing some of the channel estimation accuracy, the communication system is guaranteed to reserve the information transmission time to work properly, even the vehicle is moving at a high speed.
\vspace{-3mm}
\subsection{Large Timescale Channel Estimation}
With the small timescale channel estimation, we can obtain the CSI and optimize the hybrid beam to maximize the performance of the communication system.
Since the channel estimation method is for high-mobility communication, the real-time performance of the algorithm should be considered especially. 
Consequently, we exploit the statistical properties of channel and noise to estimate the quasi-static LoS channel for further complexity reduction.
To be specific, we can reformulate (\ref{pilotY}) with the property $\left(\bm{A}+\bm{B}\right)\odot C=\left(\bm{A}\odot C\right)+\left(\bm{B}\odot C\right)$ as 
\begin{align}
\bm{\tilde{Y}}=&\bm{\Xi}^T\left(\left(\bm{HX}\right)^T\odot\bm{G}\right)^T+\bm{\tilde{N}}\notag\\
=&\bm{\Xi}^T{\left(\left(\bm{H}_{N\!LoS}\bm{X}+\bm{H}_{LoS}\bm{X}\right)^T\odot\bm{G}\right)^T}+\bm{\tilde{N}}\notag\\
=&\bm{\Xi}^T\underbrace{\left(\left(\bm{H}_{LoS}\bm{X}\right)^T\odot\bm{G}\right)^T}_{LoS \ Channel}\notag\\
+&\bm{\Xi}^T\underbrace{\left(\left(\bm{H}_{N\!LoS}\bm{X}\right)^T\odot\bm{G}\right)^T}_{N\!LoS \ Channel}+\bm{\tilde{N}}.
\end{align}
The coherence time of the LoS channel and the NLoS channel satisfy $T^{coh}_{LoS}\gg T^{coh}_{N\!LoS}$.
We define the aggregated channel of the LoS channel and the NLoS channel in the $I$-th block as  $\bm{\bar{H}}^{LoS}_{agg}$ and $\bm{\bar{H}}^{N\!LoS}_{agg}$, which are both divided into $I$ groups.
In the $t_{coh}$-th ($t_{coh}\in\mathbb{Z}^{+}$) coherence time of NLoS channel, we can get the received signal in $I$-th block as follows
\begin{align}
\bm{\tilde{Y}}^I\left[t_{coh}\right]=\bm{{\Psi}}^I\bm{\bar{H}}_{agg}^{LoS}+\bm{{E}}^I\left[t_{coh}\right],
\end{align}
where $\bm{{E}}^I\left[t_{coh}\right]\in\mathbb{C}^{I\times N_tN_r}$ is the error matrix and defined as $\bm{{E}}^I\left[t_{coh}\right]\triangleq\bm{{\Psi}}^I\bm{\bar{H}}_{agg}^{N\!LoS}\left[t_{coh}\right]+\bm{\tilde{N}}^I\left[t_{coh}\right]$.
The index $t_{coh}$ requires to satisfy $t_{coh}\cdot T^{coh}_{N\!LoS}\leq T^{coh}_{LoS}$.
By summing up all the signal matrices, we can get the estimated LoS channel.
It is noted that when $t_{coh}=1$, we can get the estimated LoS aggregated channel as
\begin{align}
\bm{\hat{\bar{H}}}_{agg}^{LoS}&=\left(\bm{{\Psi}}^I\right)^{-1}\bm{\tilde{Y}}^I\left[t_{coh}\right]\\
&=\bm{\bar{H}}_{agg}^{LoS}+\underbrace{\bm{\bar{H}}_{agg}^{N\!LoS}\left[1\right]+\left(\bm{{\Psi}}^I\right)^{-1}\bm{\tilde{N}}^I\left[1\right]}_{error},\notag
\end{align}
in which the estimation error may be significant.
The proposed large timescale channel estimation scheme can still maintain favorable performance, provided that the Rician factor $\mathcal{K}$ and the transmission power are sufficiently high to mitigate the impact of such errors.
Notably, the sum of independent CSCG random variables remains a CSCG random variable \cite{sumcscg}. 
As the coherence time index $t_{coh}$ increases, the accuracy of the estimated LoS channel component improves progressively. 
This enables a more refined characterization of the cascaded LoS channel. 
Leveraging the estimated LoS component of the aggregated CSI $\bm{\hat{\bar{H}}}_{agg}^{LoS}$, we can pre-optimize the phase shifts of RIS.
Such a pre-optimization effectively reduces the required number of iterations, thereby lowering the computational complexity of the hybrid beamforming optimization algorithm.
\vspace{-3mm}
\section{Hybrid Beamforming Optimization with the Estimated Channels}
\vspace{-2mm}
In this section, we utilize the estimated aggregated CSI to optimize the hybrid beams, including the active beamforming at the BS side and the passive beamforming at the RIS side, to improve the system performance. 
\vspace{-4mm}
\subsection{Single VUE in Narrowband System}
We first consider a single-VUE narrowband MIMO communication system aided by a transparent RIS.
Let $\bm{x}\in\mathbb{C}^{N_t\times 1}$ denote the transmitted signal vector in the information transmission phase, which satisfies $\mathbb{E}[\|\bm{x}\|^2]=1$.
By using the active beamforming $\bm{F} \in\mathbb{C}^{N_t\times N_t}$, the BS can transmit its data vector to the VUE.
Then, from (\ref{pilotY2}), by following the groups of passive elements in $I$ blocks in Section III, the received signal vector at the VUE is given by
\vspace{-2mm}
\begin{align}
\bar{\bm{y}}=\bar{\bm{\theta}}^I\left(\left(\bar{\bm{H}}^I\right)^T\odot\bar{\bm{G}}^I\right)^T\left(\bm{Fx}\otimes\bm{I}_{N_r}\right)+\bar{\bm{n}},\label{theta_channel}
\end{align}
where $\bar{\bm{\theta}}^I=[\bar{{\theta}}_1,\bar{{\theta}}_2,\dots,\bar{{\theta}}_I]\in\mathbb{C}^{1\times I}$, $\bm{\bar{H}}^I\in\mathbb{C}^{I\times N_t}$ denote the aggregated channels grouped by passive elements. $\bm{\bar{G}}^I\in\mathbb{C}^{N_r\times I}$ can be presented as 
\begin{align} \bm{\bar{G}}^I_{(:,i)}=\sum^{l\in\mathcal{M}^I_i}_{l}\bm{G}_{(:,l)}, \ \bm{\bar{H}}^I_{(i,:)}=\sum^{l\in\mathcal{M}^I_i}_{l}\bm{H}_{(l,:)}.
\end{align}
Therefore, the estimated aggregated channel $\hat{\bm{\bar{H}}}_{agg}^I$ is an estimated version of $\left(\left(\bar{\bm{H}}^I\right)^T\odot\bar{\bm{G}}^I\right)^T$.
Then, we define $\bar{\bm{H}}_{\text{e}}^I\in\mathbb{C}^{I\times N_rN_t}\triangleq \left(\left(\bar{\bm{H}}^I\right)^T\odot\bar{\bm{G}}^I\right)^T\left(\bm{F}\otimes \bm{I}_{N_r}\right)$ as equivalent channels and $\bar{\bm{h}}_{\text{e}}^i\in\mathbb{C}^{1\times N_rN_t}$ as the $i$-th row of $\bar{\bm{H}}_{\text{e}}^I$.
$\tilde{\bm{H}}_{\text{e}}^i\triangleq \text{reshape}\left(\bar{\bm{h}}_{\text{e}}^i, N_r, N_t\right)$.

\emph{Theorem 1}: By representing $\left(\sum_{i=1}^I\bar{\theta}_i\tilde{\bm{H}}_{\text{e}}^i\right)$ as $\bm{\mathcal{H}}$, the MIMO channel data rate aided by a RIS can be given by
\begin{align}
R^{\text{nb}}=\log_2\det\left(\bm{I}_{N_r}+\frac{\bm{\mathcal{H}}\bm{\mathcal{H}}^H}{\sigma^2}\right).\label{Capacity2}
\end{align}

\begin{proof}
See Appendix A.
\end{proof}

From the formulation (\ref{Capacity2}), the data rate is determined by the phase shifts of passive elements $\bar{\bm{\theta}}^I$ and the active beamforming $\bm{F}$.
We can maximize the data rate of the vehicular MIMO system aided by the transparent RIS by jointly optimizing the active beamforming matrix $\bm{F}$ and the passive beamforming matrix $\bar{\bm{\theta}}^I$.
The data rate maximization problem can be formulated as 
\begin{subequations}
\begin{align}
\mathcal{P}0: &\max_{\bm{F},\bar{\bm{\theta}}^I}\ R^{\text{nb}}\label{P1a}\\
\text{s.t.}\quad &|\bar{\theta}_i|=1,\  i=1,\dots,I,\label{P1b}\\
&\|\bm{F}\|_{F}^2\leq P_t.\label{P1c}
\end{align}
\end{subequations}
Due to the coupling effect between the active beamforming $\bm{F}$ and the passive beamforming $\bar{\bm{\theta}}^I$, this optimization problem is difficult to solve. 
Furthermore, the phase shift constraint in (\ref{P1b}) further aggravates the challenge. 
(\ref{P1c}) is the transmit power constraint at the BS.

In the following, we give a low-complexity alternating optimization algorithm by decomposing $\mathcal{P}0$ into two sub-problems.
We can solve two sub-problems during the alternating optimization, which aim to optimize the active beamforming matrix $\bm{F}$ with given $\{\theta_m\}_
{m=1}^{M}$ or a reflection coefficient $\theta_m$ with given $\{\theta_i, i\neq m\}_
{i=1}^{M}\cup \bm{F}$.
\subsubsection{Optimization of $\bm{F}$}

There have been many algorithms to solve the problem, such as the classic water-filling strategy.
The optimal $\bm{F}$ is given by $\bm{F}^\star(\bm{F}^\star)^{H}=\bm{V}\text{diag}\left\{p_1^\star,\dots,p_{R_1}^\star\right\}\bm{V}^{H}$, where $p_r^\star$ denotes the optimal amount of power allocated to the $r$-th data stream following the water-filling strategy, and $p_r^{\star}=\max(1/p_0-\sigma^2/[\bm{\Lambda}]_{r,r},0)$, $r=1,\dots,R_1,$ with $p_0$ satisfying $\sum^{R_1}_{r=1}p_r^{\star}=P_t$. 

\subsubsection{Optimization of $\bar{{\theta}}_i$}
Based on the adaptive channel estimation method in Section III, we can get the estimated version of $\left(\left(\bar{\bm{H}}^I\right)^T\odot\bar{\bm{G}}^I\right)^T$ that presented as $\hat{\bm{\bar{H}}}_{agg}^I$.
Then, we use the estimated aggregated channel $\hat{\bm{\bar{H}}}_{agg}^I$ to optimize the group of phase shifts.
We aim to obtain the optimal $\bar{\theta}_i$ with given $\bm{F}$ and $\{\bar{\theta}_n, n\neq i\}_{i=1}^{I}$.
Then, by defining $\tilde{\bm{H}}_{-i}\triangleq\sum_{n=1, n\neq i}^I\bar{\theta}_n\tilde{\bm{H}}_{\text{e}}^n$ we can rewrite $\bm{\mathcal{H}}\bm{\mathcal{H}}^H$ as
\begin{align}
&\bm{\mathcal{H}}\bm{\mathcal{H}}^H={\left(\bar{\theta}_i\tilde{\bm{H}}_{\text{e}}^i+\tilde{\bm{H}}_{-i}\right)\left(\bar{\theta}_i\tilde{\bm{H}}_{\text{e}}^i+\tilde{\bm{H}}_{-i}\right)^H}\\
=&\tilde{\bm{H}}_{\text{e}}^i\left(\tilde{\bm{H}}_{\text{e}}^i\right)^H+\tilde{\bm{H}}_{-i}\tilde{\bm{H}}_{-i}^H+\bar{\theta}_i\tilde{\bm{H}}_{\text{e}}^i\tilde{\bm{H}}_{-i}^H+\bar{\theta}^*_i\tilde{\bm{H}}_{-i}\left(\tilde{\bm{H}}_{\text{e}}^i\right)^H\notag
\end{align}
For ease of exposition, we rewrite the objective function of $\mathcal{P}0$ in the following form with respect to $\bar{\theta}_i$:
\begin{align}
R^{\text{nb}}(\bar{\theta}_i)=\log_2\det\left(\bm{A}_i+\bar{\theta}_i\bm{B}_i+\bar{\theta}^*_i\bm{B}^H_i\right),\label{Cnb2}
\end{align}
where 
\begin{align}
&\bm{A}_i\triangleq\bm{I}_{N_r}+\left(\tilde{\bm{H}}_{\text{e}}^i\left(\tilde{\bm{H}}_{\text{e}}^i\right)^H+\tilde{\bm{H}}_{-i}\tilde{\bm{H}}_{-i}^H\right)/\sigma^2,\\
&\bm{B}_i\triangleq\tilde{\bm{H}}_{\text{e}}^i\tilde{\bm{H}}_{-i}^H/\sigma^2.\label{Bi}
\end{align}

\emph{Case I: $|\mathcal{M}^I_i|=$1:} This means there is only one passive element in $i$-th group.
In this case, the closed form solution of the optimal $\bar{\theta}_i^{\star}$ has been proved as follows \cite{9110912}
\begin{align}
\bar{\theta}_i^{\star}=
    \begin{cases}
    e^{-j\arg\{\lambda_i\}},  \  &\text{if} \  \text{tr}\left(\bm{A}^{-1}_i\bm{B}_i\right)\neq 0,\\
    1, &\text{otherwise},
    \end{cases}
    \label{optimaltheta}
\end{align}
where $\lambda_i$ denotes the sole non-zero eigenvalue of $\bm{A}^{-1}_i\bm{B}_i$.

\emph{Case II: $|\mathcal{M}^I_i|>$1:} There is more than one passive element in $i$-th group.
Based on \emph{Case I}, the rank of matrix $\bm{A}^{-1}_i\bm{B}_i$ is bigger than one, for which we have the following lemma.

\emph{Lemma 1:} If $N_r>2$, the general closed form solution of optimal $\bar{\theta}_i$ for (\ref{Cnb2}) using only elementary operations and radicals does not exist.

\begin{proof}
See Appendix C.
\end{proof}

For the case that $N_r>2$, since there is no closed-form solution, we can optimize the phases of passive elements group by group with the classical gradient ascent method, and at least one local optimal solution can be obtained.
To start with, the derivative of $R^{\text{nb}}$ in (\ref{Cnb2}) with respect to $\bar{\phi}_i$ can be given as 
\begin{align}
d(\bar{\phi}_i)\triangleq\frac{\partial R^{\text{nb}}}{\partial\bar{\phi}_i}&=\frac{j}{\text{ln}2}\text{tr}\left(\textbf{X}_i^{-1}\left(\bar{\theta}_i\bm{B}_i-\bar{\theta}_i^*\bm{B}_i^H\right)\right),
\end{align}
where $\textbf{X}_i\triangleq \bm{A}_i+\bar{\theta}_i\bm{B}_i+\bar{\theta}^*_i\bm{B}^H_i$ and $\bar{\phi}_i$ is the $i$-th group phase that satisfies $\bar{\theta}_i=e^{j\bar{\phi}_i}$.  
By initializing $\bar{\phi}_i^{(0)}$, we can obtain $\bar{\phi}_i^{(t)}=\bar{\phi}_i^{(t-1)}+\beta d(\bar{\phi}^{(t-1)}_i)$ with $\beta$ defined as the learning rate.
After repeated iterations we can obtain the optimal $\bar{\phi}_i^{\star}$ and the optimal $\bar{\theta}_i^{\star}=e^{j\bar{\phi}_i^{\star}}$.

However, in the practical algorithm design of the proposed system, the objective is to determine the optimal phase shifts of passive elements to maximize a predefined objective function (\ref{Cnb2}).
Exhaustive search method often outperforms gradient ascent method, particularly when the optimization variables such as phase shifts are quantized. 
Gradient ascent relies on the assumption that the search space is continuous and differentiable, which is incompatible with the discrete nature of quantized phase values in practical systems. 
Moreover, the existence of multiple local optima and the non-convexity of the objective function can significantly degrade the performance of gradient-based approaches, which are typically sensitive to initialization and prone to premature convergence. 
In contrast, exhaustive search guarantees identification of the global optimum within the quantized domain and offers greater robustness under these conditions. 
More importantly, when the passive elements are grouped, the dimensionality of the search space is further reduced, making exhaustive search computationally feasible. 
With optimal solutions to the active beamforming and passive beamforming,
we are ready to complete our proposed alternating optimization algorithm for solving $\mathcal{P}\text{0}$.
Specifically, we first randomly generate the sets of $\{\bar{{\theta}}_i\}_{i=1}^I$.
Then, by obtaining the optimal active beamforming matrix $\bm{F}$ and $\{\bar{{\theta}}^I_i\}_{i=1}^I$ according to the large time scale estimated channel $\hat{\bar{\bm{H}}}_{agg}^{LoS}$, the algorithm then proceeds by iteratively optimizing $\bm{F}$ and $\bar{\bm{\theta}}$ according to the small time scale estimated channel $\hat{\bar{\bm{H}}}_{agg}$ until convergence is reached.
\subsection{Multi VUEs in Broadband System}
We solved $\mathcal{P}$0 in the simple scenario of a single VUE in narrowband system.
However, the scenario of broadband communication system with multiple VUEs are more realistic.
In this subsection, we extend our algorithm in the narrowband MIMO system to the broadband MIMO-OFDM system in the high mobility scenario.
Similarly, we consider a MIMO communication system with $N_t$ antennas at the BS and $N_r$ antennas at the VUE. 
Each VUE is equipped with one transparent RIS at the top of the vehicle, where $M$ passive elements are deployed.
Due to the multiplicative fading between any two RISs, signals that have been reflected twice or more can be ignored.
Assuming that with the two-timescale channel estimation method, the CSI of each sub-carrier can be obtained.

Let $N,K$ and $\Delta f$ denote the number of sub-carriers, the number of VUEs, and the sub-carrier interval, respectively, and let $\bm{\bar{H}}[k,n]\in\mathbb{C}^{I\times N_t}$, $\bm{\bar{G}}[k,n]\in\mathbb{C}^{N_r\times I}$, $k\in\mathcal{K}$, $n\in\mathcal{N}$ denote the channel matrices of $k$-th VUE in $n$-th sub-carrier, respectively, where $\mathcal{K}\triangleq\{1,\dots,K\}$, $\mathcal{N}\triangleq\{1,\dots,N\}$.
Similar to our treatment of narrowband single-VUE scenarios, let $\bar{\bm{H}}_{\text{e}}[k,n]\triangleq \left(\left(\bar{\bm{H}}[k,n]\right)^T\odot\bar{\bm{G}}[k,n]\right)^T\left(\bm{F}[n]\otimes \bm{I}_{N_r}\right)$ denote the estimated equivalent channel,   
$\bar{\bm{h}}_{\text{e}}^i[k,n]$ is the $i$-th row of $\bar{\bm{H}}_{\text{e}}[k,n]$ and 
$\tilde{\bm{H}}_{\text{e}}^i[k,n]\triangleq \text{reshape}\left(\bar{\bm{h}}_{\text{e}}^i[k,n], N_r, N_t\right)$.
Note that due to the lack of baseband processing capabilities, all phase shift matrices of $K$ RISs are the same in $N$ sub-carriers.
Owing to different mobility-aware channel conditions and the non-orthogonality induced by multiple sub-carriers, our proposed communication system has to overcome the performance degradation induced by ICI from Doppler spread effect.
From \cite{ICI}, we can reformulate the ICI of the $k$-th VUE in $n$-th sub-carrier as the VUE-self ICI and VUE-inter ICI as shown in formulation (\ref{ICI}),
\begin{align}
ICI_{k,n}=&\frac{1}{2\Delta f^2}\Bigg({\vphantom{\sum^{K}_{\substack{d=1 \\ d\neq k}}\sum^{l\in\mathcal{N}_d}_{l}\left(\frac{f_c}{c}v_d\right)^2\frac{p_l}{\left(l-n\right)^2}}\sum^{l\in\mathcal{N}_k}_{l\neq n}\left(\frac{f_c}{c}v_k\right)^2\frac{p_{k,n}}{\left(l-n\right)^2}}\notag\\
&+{\vphantom{\sum^{K}_{\substack{d=1 \\ d\neq k}}\sum^{l\in\mathcal{N}_d}_{l}\left(\frac{f_c}{c}v_d\right)^2\frac{p_{d,l}}{\left(l-n\right)^2}}\sum^{K}_{\substack{d=1 \\ d\neq k}}\sum^{l\in\mathcal{N}_d}_{l}\left(\frac{f_c}{c}v_d\right)^2\frac{p_{d,l}}{\left(l-n\right)^2}}\Bigg),\label{ICI}
\end{align}
where $f_c$, $c$, $v_k$, $\mathcal{N}_k$ and $p_{k,n}$ denote the operating carrier frequency, the speed of light, velocity of $k$-th VUE, sub-carrier index set allocated to $k$-th VUE and the power allocated to $n$-th sub-carrier $k$-th VUE, respectively.
In order to simplify the exposition of the performance optimization problem in MIMO-OFDM system, we consider a binary indicator $\rho_{k,n} \in \{0, 1\}$ where $\rho_{k,n} = 1$ implies that the $n$-th sub-carrier is allocated to $k$-th VUE and $\rho_{k,n} = 0$ is not.
It is worth noting that for each sub-carrier, it can only be allocated to one VUE, which means $\sum_{k=1}^K\rho_{k,n}=1$.
With the introduced binary indicator, the ICI can be reformulated as 
\begin{align}
ICI_{k,n}=\frac{1}{2}\left(\frac{f_c}{\Delta f\cdot c}\right)^2\sum_{d=1}^{K}\sum_{l\neq n}^{l\in\mathcal{N}_d}v_d^2\frac{\rho_{d,l}p_{d,l}}{\left(l-n\right)^2}.
\end{align}
The active beamforming requires to be designed for each sub-carrier and we normalize the power of the active transmit matrix in all sub-carriers, which means in the $n$-th sub-carrier, the individual $\bm{F}[n]$ satisfies $\|\bm{F}[n]\|_F^2=1$, $n\in\mathcal{N}$.
Considering the influence of ICI and representing $\left(\sum_{i=1}^I{\theta}_i[k]\tilde{\bm{H}}_{\text{e}}^i[k,n]\right)$ as $\bm{\mathcal{H}}[k,n]$, the data rate of the proposed high-mobility MIMO-OFDM system is thus given by
\begin{align}
C^{\text{bb}}_{k,n}=&\Delta f\log_2\left|\bm{I}_{N_r}+\rho_{k,n}p_{k,n}\frac{\bm{\mathcal{H}}[k,n]\bm{\mathcal{H}}[k,n]^H}{ICI_{k,n}+N_0\cdot \Delta f}\right|.\label{Cwb}
\end{align}
We can formulate the data rate maximization problem at the end of this section.
Similar to $\mathcal{P}$0, (\ref{p2b})-(\ref{p2d}) specify the limitation of phase shifts and the normalized transmit power constraint of active beamforming;
(\ref{p2e}) limits the maximum power per carrier to avoid a situation where a well-conditioned carrier consumes all the power and destroys fairness;
(\ref{p2f}) indicates that the total allocated power constraint in all sub-carriers; 
(\ref{p2g}) guarantees the minimal quality of service (QoS) requirements of $C_{\min}$ for all VUEs; 
Finally, constraints (\ref{p2h}) and (\ref{p2i}) confine the value of the binary indicator $\rho_{k,n}$, which have been discussed before.
In Problem $\mathcal{P}1$, we have a total of four groups of optimization variables, namely, the sub-carrier allocated binary indicator, transmit power allocation, active beamforming, and phase shifts of all RISs. 
Problem $\mathcal{P}1$ is regarded as a non-convex mixed-integer non-linear problem (MINLP) and hence an NP-hard problem. 
Therefore, it is significantly challenging to obtain a globally optimal solution directly due to the coupling effect of all variables.
In our work, a locally optimal solution is provided.
Specifically, the optimization of the former two variables is related to resource allocation, while the optimization of remaining variables specifies the hybrid beamforming design.
Thus, we decouple $\mathcal{P}1$ into resource allocation problem and beamforming optimization problem. 
By optimizing subproblems one by one alternatively with the other settings fixed, at least a locally optimal solution at low complexity can be achieved.
In the following, we will give the specific algorithm.
\begin{subequations}
\begin{align}
\mathcal{P}1:\ &\max_{\substack{\rho_{k,n},p_{k,n},\bm{\Phi}^I[k],\bm{F}[n]}} \sum_{k=1}^{K}\sum_{n=1}^NC^{\text{bb}}_{k,n}\label{p2a}\\
\text{s.t.}\quad &\bm{\Phi}^I[k]=\text{diag}\left\{\theta_1[k],\dots,\theta_I[k]\right\}\label{p2b},\\
&|\theta_i[k]|=1,\  i=1,\dots,I,\label{p2c}\\
&\|\bm{F}[n]\|_{F}^2\leq 1,\label{p2d}\\
&0\leq p_{k,n}\leq P_{\text{max}},\label{p2e}\\
&\sum_{k=1}^K\sum_{n=1}^Np_{k,n}\leq P_{\text{tot}},\label{p2f}\\
&\sum_{n=1}^NC^{\text{bb}}_{k,n}\geq C_{\min}, \ k\in\mathcal{K},\label{p2g}\\
&\rho_{k,n}\in\{0,1\}, \  k\in\mathcal{K},\label{p2h}\\
&\sum_{k=1}^K\rho_{k,n}=1\label{p2i}.
\end{align}
\end{subequations}

\subsubsection{Optimization of Resource Allocation While Fixing the
Beamforming Settings}
Given the active beamforming matrix $\bm{F}$ and the RIS phase shift vector $\bm{\theta}$, Problem $\mathcal{P}1$ can be simplified to
\begin{align}
\mathcal{P}2:\ &\max_{\substack{\rho_{k,n},p_{k,n}}} \sum_{k=1}^{K}\sum_{n=1}^NC^{\text{bb}}_{k,n}\\
&\text{s.t.}\quad (\text{\ref{p2e})-(\ref{p2i}}).\notag
\end{align}
For the problem $\mathcal{P}2$, we can first observe that in (\ref{Cwb}), the continuous term $p_{k,n}$ and discrete term $\rho_{k,n}$ are tough to handle.
To deal with this, we define an auxiliary parameter denoted by $\widehat{p}_{k,n}$
as
\begin{align}
\widehat{p}_{k,n}=
    \begin{cases}
    0,  \  &\text{if} \  \rho_{k,n}= 0,\\
    \rho_{k,n},p_{k,n}, &\text{otherwise}.
    \end{cases}
\end{align}
The new constraints related to power constraints (\ref{p2e}) and (\ref{p2f})
 can be equivalently introduced as
\begin{align}
&0\leq \widehat{p}_{k,n}\leq P_{\text{max}},\label{hatp1}\\
&\sum_{k=1}^K\sum_{n=1}^N\widehat{p}_{k,n}\leq P_{\text{tot}}.\label{hatp2}
\end{align}
Therefore, we can reformulate the Problem $\mathcal{P}2$ by replacing the auxiliary parameter as 
\begin{align}
\mathcal{P}\text{2-}E:\ &\max_{\substack{\widehat{p}_{k,n}}} \sum_{k=1}^{K}\sum_{n=1}^NC^{\text{bb}}_{k,n}\\
&\text{s.t.}\quad (\ref{hatp1}),(\ref{hatp2}),(\text{\ref{p2g})-(\ref{p2i}}).\notag
\end{align}
Next, to deal with the binary variables problem, we apply the integer relaxation of binary variables strategy to relax the sub-carrier assignment indicator within the region between 0 and 1, which can be expressed as
\begin{align}
\widehat{\bm{\rho}}\triangleq\left\{0\leq \widehat{\rho}_{k,n}\leq 1 | \forall n\in\mathcal{N},k\in\mathcal{K}\right\}.\label{IntRel}
\end{align}
Inspired by \cite{IntRel}, to constrain the continuous relaxation variables $\widehat{\rho}_{k,n}$ to 0 and 1, we design a penalty function as follows
\begin{align}
G(\widehat{\bm{\rho}})=\sum_{k=1}^K\sum_{n=1}^N\widehat{\rho}_{k,n}\left(1-\widehat{\rho}_{k,n}\right)\leq 0.\label{Gp}
\end{align}
It is obvious that $\mathcal{P}\text{2-}E$ stays unchanged with the employment of integer relaxation under constraint (\ref{Gp}) since combining (\ref{IntRel}), (\ref{Gp}) yields $\widehat{\rho}_{k,n}\in\{0,1\}$.
However, the new introduced constraint (\ref{Gp}) is non-convex.
Consequently, we employ the abstract Lagrangian duality \cite{aLd} to solve it.
We construct the Lagrangian objective which can be given as
\begin{align}
\Upsilon\left(\widehat{\bm{p}},\widehat{\bm{\rho}},\lambda\right)=\sum_{k=1}^{K}\sum_{n=1}^NC^{\text{bb}}_{k,n}-\lambda G(\widehat{\bm{\rho}}),\label{upsilon}
\end{align}
where $\lambda$ is the penalty constant.
The transformed problem can be reformulated as a primal problem as
\begin{align}
\mathcal{P}\text{2-}E1:\ &\max_{\{\widehat{\bm{p}},\widehat{\bm{\rho}}\}\in\mathcal{D}} \min_{\lambda} \ \Upsilon\left(\widehat{\bm{p}},\widehat{\bm{\rho}},\lambda\right),
\end{align}
where $\mathcal{D}$ is defined as the set of feasible solutions that satisfies all constraints (\ref{hatp1}),(\ref{hatp2}),(\text{\ref{p2g})-(\ref{p2i}}).
Note that the problem $\mathcal{P}\text{2-}E1$ is in the form of max-min with $\lambda$ existing in the minimization operator. 
This is because minimization of the negative part of the penalty function in (\ref{upsilon}) aims to sustain the constraints and satisfy the non-penalized problem in (41).
The transformed problem $\mathcal{P}\text{2-}E1$ is equivalent to the unrelaxed problem $\mathcal{P}\text{2-}E$ with the attained sufficiently large value of $\lambda$.


By now, we have achieved Problem $\mathcal{P}\text{2-}E1$.
However, the objective function $\Upsilon\left(\widehat{\bm{p}},\widehat{\bm{\rho}},\lambda\right)$ is a non-concave form.
In order to facilitate the problem, we transform the objective function into the subtraction of two concave functions, which can be presented as
\begin{align}
\Upsilon\left(\widehat{\bm{p}},\widehat{\bm{\rho}},\lambda\right)=\sum_{k=1}^KF^k_1\left(\widehat{\bm{p}},\widehat{\bm{\rho}}\right)-\sum_{k=1}^KF^k_2\left(\widehat{\bm{p}},\widehat{\bm{\rho}}\right)-\lambda G(\widehat{\bm{\rho}}),\label{F1MF2}
\end{align}
where functions $F^k_1$, $F^k_2$ are defined as follows
\vspace{-2mm}
\begin{align}
&F^k_1\triangleq\sum_{n=1}^N\log_2\det\Big(\left(\widehat{ICI}_{k,n}+N_0\cdot \Delta f\right)\bm{I}_{N_r}\notag\\
&\quad\quad\quad\quad\quad\quad\quad\quad\quad\quad\quad+\widehat{p}_{k,n}\bm{\mathcal{H}}[k,n]\bm{\mathcal{H}}[k,n]^H\Big),\\
&F^k_2\triangleq\sum_{n=1}^NN_r\log_2\left(\widehat{ICI}_{k,n}+N_0\cdot \Delta f\right).\label{F2}
\end{align}
$F^k_1$ and $F^k_2$ are both concave functions with respect to $\widehat{\bm{p}}$. 
To calculate the optimal $\widehat{\bm{p}}$ under all constraints, we use the D.C. algorithm to handle this non-concave problem.
By employing first-order Taylor approximation on $F^k_2\left(\widehat{\bm{p}},\widehat{\bm{\rho}}\right)$, the concave function can be transformed into affine one so that we can use existing convex optimization solvers. 
We first approximate the term $F^k_2(\widehat{\bm{p}})$ as  
\begin{align}
\tilde{F}^k_2(\widehat{\bm{p}}[j])=&F^k_2\left(\widehat{\bm{p}}[j-1]\right)\notag\\
&+\triangledown^T_{\widehat{\bm{p}}} F^k_2\left(\widehat{\bm{p}}[j-1]\right)\cdot\left(\widehat{\bm{p}}[j]-\widehat{\bm{p}}[j-1]\right),\label{waveF2}
\end{align}
where $\triangledown^T_{\widehat{\bm{p}}}F^k_2\left(\widehat{\bm{p}}\right)$ is the first-order partial derivatives of $F^k_2\left(\widehat{\bm{p}}\right)$ with respect to $\widehat{\bm{p}}$.
The index $j$ means the iteration count.
Then, the term $F^k_2\left(\widehat{\bm{p}},\widehat{\bm{\rho}}\right)$ can be approximated
as
\vspace{-3mm}
\begin{align}
J\left(\widehat{\bm{p}}[j],\widehat{\bm{\rho}}[j]\right)=&\sum_{k=1}^K\tilde{F}^k_2(\widehat{\bm{p}}[j])+\lambda\bigg(G\left(\widehat{\bm{\rho}}[j-1]\right)\label{Jprho}\\
&+\triangledown^T_{\widehat{\bm{\rho}}} G\left(\widehat{\bm{\rho}}[j-1]\right)\cdot\left(\widehat{\bm{\rho}}[j]-\widehat{\bm{\rho}}[j-1]\right)\bigg),\notag
\end{align}
where $\triangledown^T_{\widehat{\bm{\rho}}}G\left(\widehat{\bm{\rho}}\right)$ is the first-order partial derivatives of  $G\left(\widehat{\bm{\rho}}\right)$ with respect to $\widehat{\bm{\rho}}$.
The partial derivatives of $F^k_2\left(\widehat{\bm{p}}\right)$ and $G\left(\widehat{\bm{\rho}}\right)$ can be further \textcolor{black}{presented by
\begin{align}
&\frac{\partial F^k_2(\widehat{\bm{p}})}{\partial \widehat{{p}}_{k,n}}\label{partialp}\\
=&\frac{N_r}{2\ln 2}\left(\frac{f_c}{\Delta f\cdot c}\right)^2\sum^K_{d=1}\sum^N_{\substack{l=1\\ l\neq n}}\frac{v_d^2}{\left(ICI_{d,l}+N_0\cdot \Delta f\right)}\frac{1}{\left(l-n\right)^2},\notag\\
&\frac{\partial G(\widehat{\bm{\rho}})}{\partial \widehat{\rho}_{k,n}}=1-2\widehat{\bm{\rho}}_{k,n}.\label{partialrho}
\end{align}
Based on (\ref{Jprho}), (\ref{partialp}) and (\ref{partialrho})}, we can approximate the objective function (\ref{F1MF2}) as 
$\Upsilon_{app}\left(\widehat{\bm{p}},\widehat{\bm{\rho}}\right)=\sum_{k=1}^{K}F^k_1\left(\widehat{\bm{p}},\widehat{\bm{\rho}}\right)-J\left(\widehat{\bm{p}},\widehat{\bm{\rho}}\right).$
We can approximate $\mathcal{P}\text{2-}E1$ and constraints into the following formulation as
\begin{subequations}
\begin{align}
\mathcal{P}\text{3}:\ &\max_{\widehat{\bm{p}},\widehat{\bm{\rho}}} \ \Upsilon_{app}\left(\widehat{\bm{p}},\widehat{\bm{\rho}}\right),\\
&\text{s.t.}\quad (\ref{hatp1}),(\ref{hatp2}),(\ref{IntRel}), (\text{\ref{p2i}}),\notag\\
&\quad\quad F^k_1(\widehat{\bm{p}})-\tilde{F}^k_2(\widehat{\bm{p}})\geq C_{\min}, \ k\in\mathcal{K}\label{P4c}.
\end{align}
\end{subequations}
The transformed problem $\mathcal{P}\text{3}$ is convex and will converge to a local optimum.
Now, Problem $\mathcal{P}\text{2}$ has been transformed into a standard concave problem and can be thus optimally solved by existing convex
optimization solvers such as CVX.
It is worth noting that since the concave property $F_2(\widehat{\bm{p}},\widehat{\bm{\rho}})\leq J\left(\widehat{\bm{p}},\widehat{\bm{\rho}}\right)$ holds, we can achieve that $\Upsilon\left(\widehat{\bm{p}},\widehat{\bm{\rho}}\right)\geq \Upsilon_{app}\left(\widehat{\bm{p}},\widehat{\bm{\rho}}\right)$ holds.
Accordingly, the first-order Taylor approximation will provide a smaller feasible
solution set for $\widehat{\bm{p}}$ and $\widehat{\bm{\rho}}$.
The constraint (\ref{P4c}) will not change the original throughput satisfaction.

\subsubsection{Optimization of Hybrid Beamforming While Fixing the
Resource Settings}
Given the sub-carrier allocation $\bm{\rho}$ and the power allocation $\bm{p}$, Problem $\mathcal{P}\text{1}$ is reformulated as
\begin{align}
\mathcal{P}4:\ &\max_{\substack{\bm{\Phi}^I[k],\bm{F}[n]}} \sum_{k=1}^{K}\sum_{n=1}^NC^{\text{bb}}_{k,n}\\
&\text{s.t.}\quad \text{(\ref{p2b})}, \text{(\ref{p2c})}, \text{(\ref{p2d})}\notag.
\end{align}
We define the sub-carrier allocation set as $\{\mathcal{N}_k|\forall k\in \mathcal{K}\}$, where $\mathcal{N}_k\triangleq\{n|\rho_{k,n}=1\}$.
Note that in our formulated problem, sub-carrier indexes assigned to other VUEs do not affect the $k$-th VUE because ICI is always existing.
Moreover, due to the severe \emph{distance-product} power loss over multiple reflections and the penetration loss of the metal panel, we ignore the inter-VUE interference.
Therefore, Problem $\mathcal{P}4$ can be reduced to maximize each VUE's channel data rate.
For example, the $k$-th VUE's data rate maximization is given by
\begin{align}
\mathcal{P}\text{4-}E:\ &\max_{\substack{\bm{\Phi}^I[k],\bm{F}[n]}} \sum^{n\in\mathcal{N}_k}C^{\text{bb}}_{k,n}\\
&\text{s.t.}\quad \text{(\ref{p2b})}, \text{(\ref{p2c})}, \text{(\ref{p2d})}.\notag
\end{align}
Given the resource allocation setting, the $ICI_{k,n}$ and noise are fixed. 
Similar to what we do when dealing with the narrowband single-VUE system, we optimize active beamforming matrice over all allocated sub-carriers with passive beamforming fixed.
In each sub-carrier, we still apply the classic water-filling strategy to obtain optimal $\bm{F}[n]$.
Similarly, the optimal $\bm{F}[n]$ is given by $\bm{F}^\star[n](\bm{F}^\star[n])^{H}=\bm{V}[n]\text{diag}\left\{p_1^\star,\dots,p_r^\star\right\}\bm{V}^{H}[n]$, where $\bm{V}[n]$ is the right singular matrix of equivalent channel $\bm{H}^I_{\text{e}}[n]$ in $n$-th sub-carrier, $r=\text{rank}(\bm{H}^I_{\text{e}}[n])$ and $p_i^\star$ denotes the optimal amount of power allocated to the $i$-th data stream following the water-filling strategy. 
The optimal liner precoding matrix $\bm{F}^\star[n]=\bm{V}[n]\text{diag}\left\{p_1^\star,\dots,p_r^\star\right\}^{\frac{1}{2}}$.

Then, we discuss how to optimize passive beamforming over all allocated sub-carriers with active beamforming fixed.
Our proposed optimization algorithm in Section IV.A in the narrowband case cannot be directly applied to solve $\mathcal{P}\text{4-}E$ since the objective function is a non-concave function over the phase shift $\{\theta_i[k]\}^I_{i=1}$.
To simplify the processing of the Problem $\mathcal{P}\text{4-}E$, we relax the constraint (\ref{p2c}) into $|\theta_i[k] \leq 1, i = 1, . . . , I$.
With the active beamforming $\{\bm{F}[n]\}^N_{n=1}$ and other $I-1$ groups reflection phase shifts $\{\theta_i,i\neq m\}^I_{i=1}$ given, the objective problem is
a concave function over the remaining group reflection coefficient $\theta_m$.
Thus, $\mathcal{P}\text{5}$ is a convex optimization problem over $\theta_m$, and the optimal solution can be obtained via CVX.
If all reflection coefficients satisfy the constraints in (\ref{p2c}) of $\mathcal{P}1$, then the relaxation is tight, and the result $\theta_i$ is also a local optimum solution for $\mathcal{P}1$.
Otherwise, an approximate solution for $\mathcal{P}1$ can be obtained by normalizing the amplitude of $\theta_i$.
The group phase shifts can be optimized group by group until convergence is reached.
Then we can alternately optimize subcarrier and power allocation, the active beamformer, and the RIS phase shifts while explicitly modeling the Doppler-induced ICI.
\vspace{-3mm}
\section{Simulation Results}
In this section, we give the performance evaluation of the adaptive channel estimation method and the hybrid beamforming algorithm through simulations.
Considering a typical scenario in which a vehicle is traveling in a straight line on a road, both the BS and VUE are equipped with a uniform linear array with the numbers of antenna are $N_t=16$ and $N_r=25$, respectively; while the RIS is equipped with a uniform planar array.
The distance from the BS to the RIS on the vehicle roof and from the RIS to the VUE are set as $d_{BR}=1500$ m and $d_{RV}=2$ m, respectively.
The distance-dependent path loss for all channels is modeled as $P_{L}=P_0(d/d_0)^{-\bar{\alpha}}$, where
$P_0$ = -30 dB denotes the path loss at the reference distance $d_0 = 1$ m; $\bar{\alpha}$ denotes the path loss exponent.
In our model, the path loss exponent of the BS-RIS link and the RIS-VUE link are set as $\bar{\alpha}_{BR}=2.2$ and $\bar{\alpha}_{RV}=2.8$, respectively.
$\mathcal{K}=5$ dB denotes the Rician factor. 
The noise power spectrum density is set as -174 dBm/Hz. 
The center frequency $f_c=3.5$ GHz and bandwidth of the narrowband MIMO system is set as $B=1$ MHz.
In the uplink channel estimation phase, the transmit power of the pilot signal is set as $P_u$ = 30 dBm.

Firstly, we evaluate the normalized mean square error (NMSE) performance of adaptive channel estimation method. 
The NMSE of channel is defined by $\text{NMSE}\triangleq \frac{||\hat{\bm{H}}-\bm{H}||^2_F}{||\bm{H}||^2_F}.$

\begin{figure}{}
\centering
  \includegraphics[width=2.8in,height=2in]{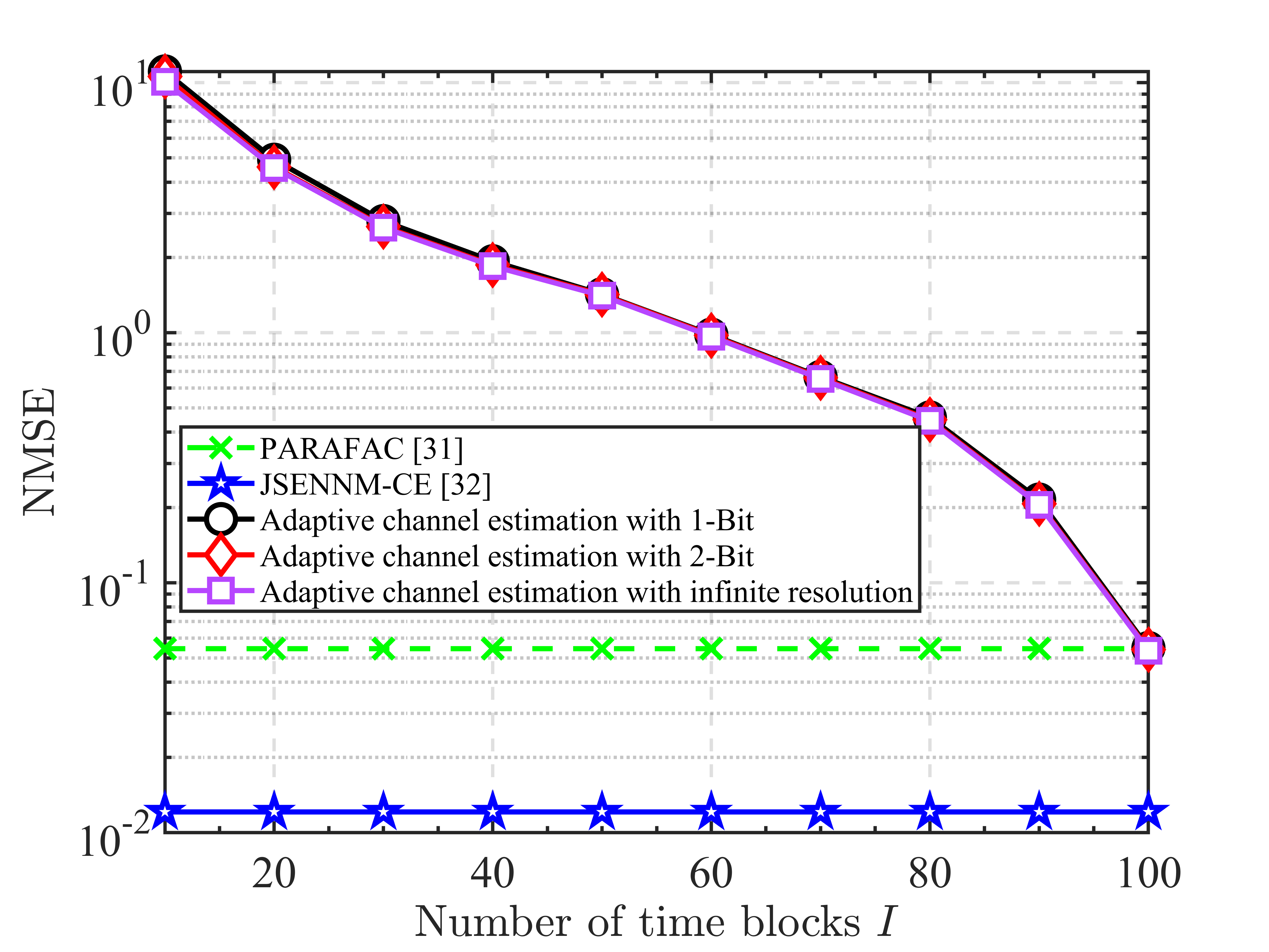}
\vspace{-4mm}
  \caption{NMSE versus number of time blocks $I$ for RIS-aided
 vehicular MIMO communication with $M =100$.}\label{fig2}
\vspace{-4mm}
\end{figure}

Fig. \ref{fig2} shows the NMSE of the cascaded channel against the number of blocks.
We consider the following two existing channel estimation schemes PARAFAC \cite{9104260} and JSENNM-CE \cite{10974631} for comparison.
We can observe that the NMSE of the adaptive channel estimation method decreases as the number of blocks increases. 
JSENNM-CE has the best performance.
And our proposed algorithm achieves the same performance as the PARAFAC when $I=M$.
This is obvious because when $I=M$, the grouping scheme of reflective elements fails and our proposed algorithm is essentially no different from the traditional channel estimation scheme.
We also compare the performance of channel estimation with different resolutions of the reflective element phase shifter.
They show similar performance because the mutual orthogonality of the pilot phases can be easily ensured in the channel estimation phase, even if the resolution of the practical phase shifters is low.


\begin{figure}{}
\centering
  \includegraphics[width=2.4in,height=2.6in]{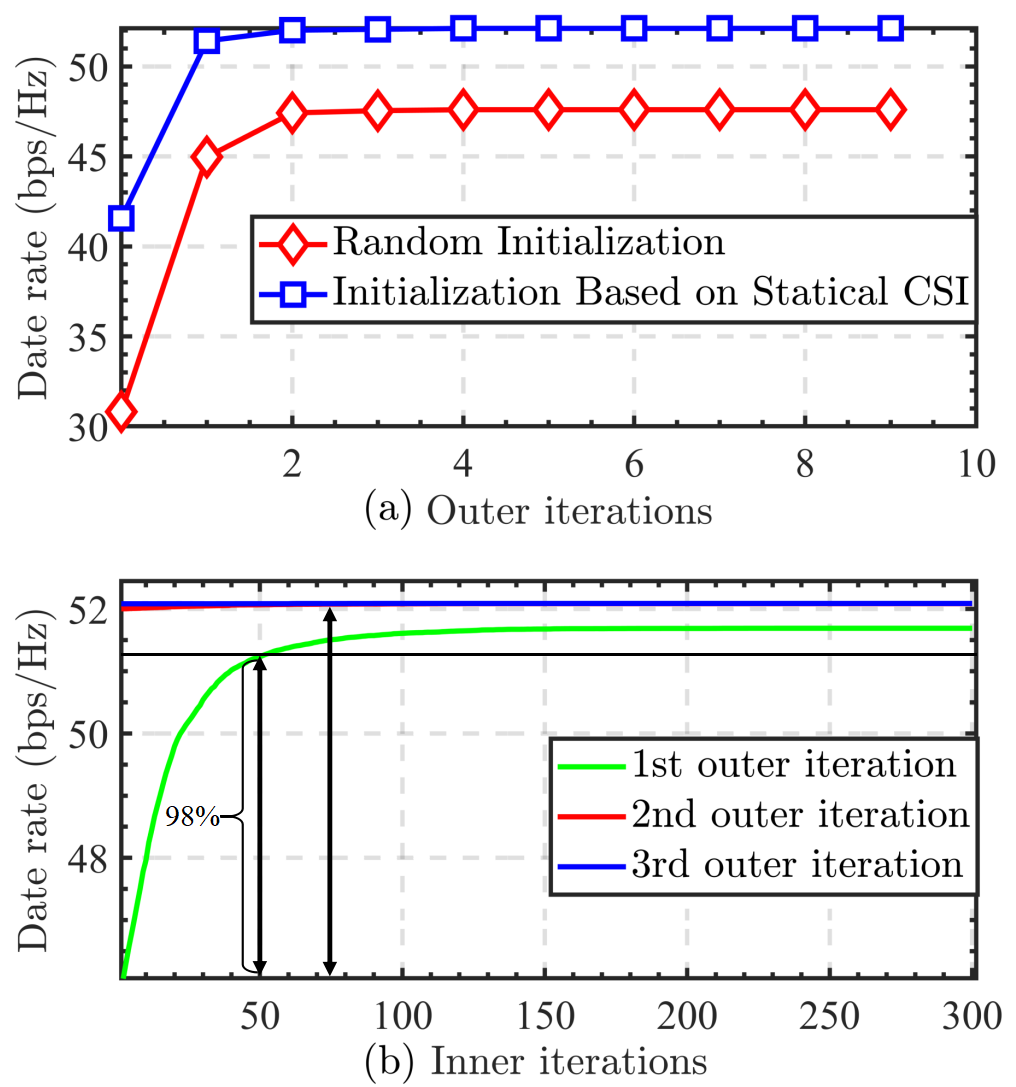}
\vspace{-4mm}
  \caption{(a) Convergency of the proposed algorithm in outer iterations. (b) Convergency of the proposed algorithm in inner iterations}\label{fig7}
\vspace{-6mm}
\end{figure}

Next, we evaluate the performance of the hybrid beamforming algorithm for single VUE in narrowband system with transmit power $p_t=20$ dBm, $M=100$ and $I=20$.
In Fig. \ref{fig7}, we show the convergence behavior of beamforming algorithm with the adaptive estimated channel.
It is observed that the proposed beamforming algorithm converges monotonically, which is consistent with our analysis in Section VI.
In outer iteration, the data rate can converge in the first two iterations.
While, in inner iteration, the convergence speed is also fast.
The average number of inner iterations in first outer iteration needed to achieve the almost convergence is 50, at which performance reaches $98\%$ of the fully converged performance.
It is also observed that the optimization that starts at initialization based on Statical CSI performs better than that starts at random initialization.
Since selecting better performing initial points means that the optimized variables are closer to the global optimum, not only do they converge faster, but they also have better performance when they finally converge.
In the following, under $p_t=20$ dBm and $M=100$, we compare the performance of proposed beamforming algorithm with the following benchmark schemes: Baseline 1, 2, 3, 4 is the data rate of (\ref{Capacity2}) obtained based on the perfect CSI with no passive elements grouping scheme, the traditional complete channel estimation  with no passive elements grouping scheme, the perfect CSI with  passive elements grouping scheme and the statical CSI with passive elements grouping scheme, respectively.

\begin{figure}{}
\centering
  \includegraphics[width=2.8in,height=2in]{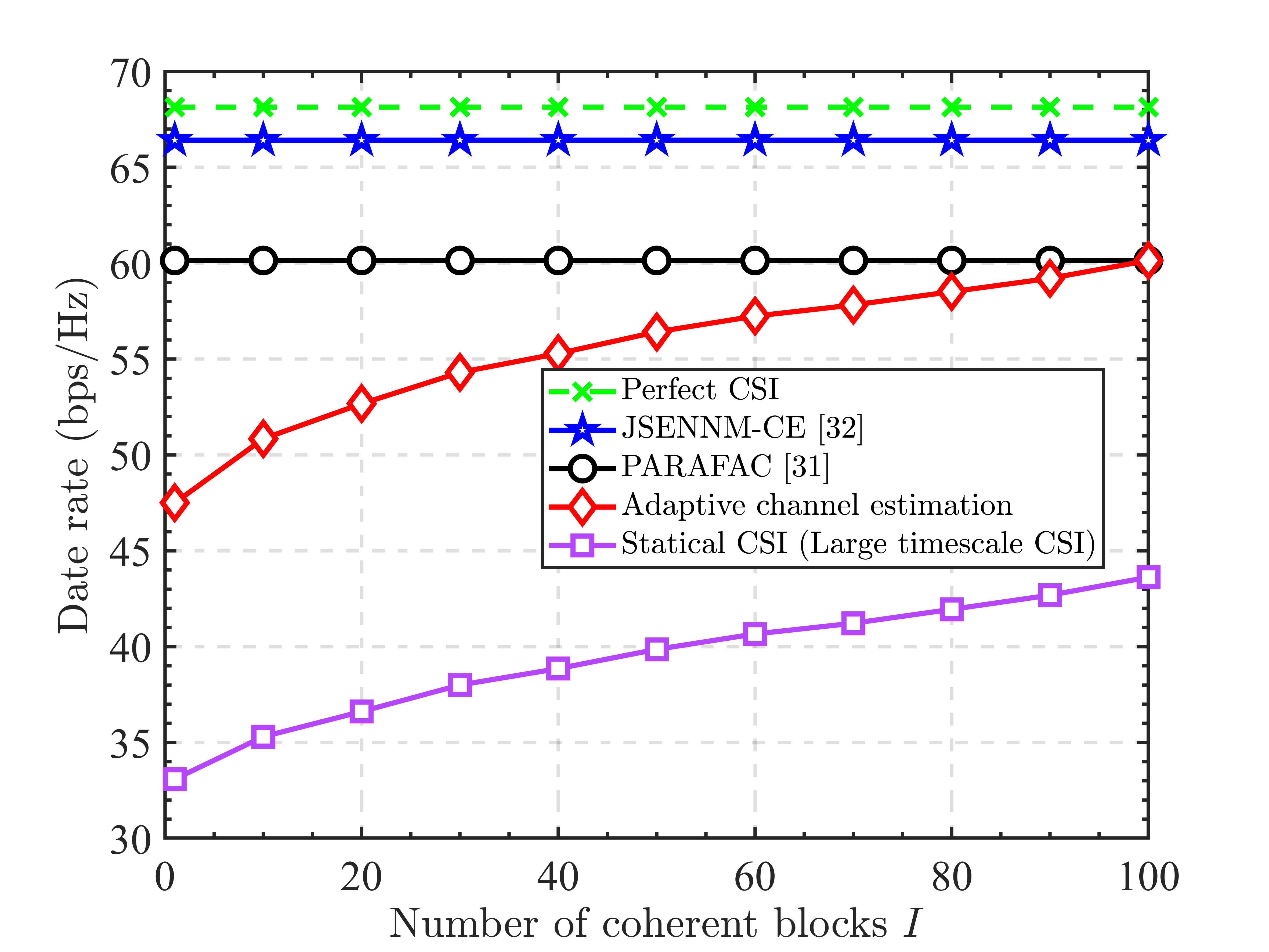}
\vspace{-4mm}
  \caption{Data rate versus number of time blocks $I$ for RIS-aided
 vehicular MIMO communication with $M=100$.}\label{fig5}
\vspace{-4mm}
\end{figure}
\begin{figure}{}
\centering
  \includegraphics[width=3in,height=2in]{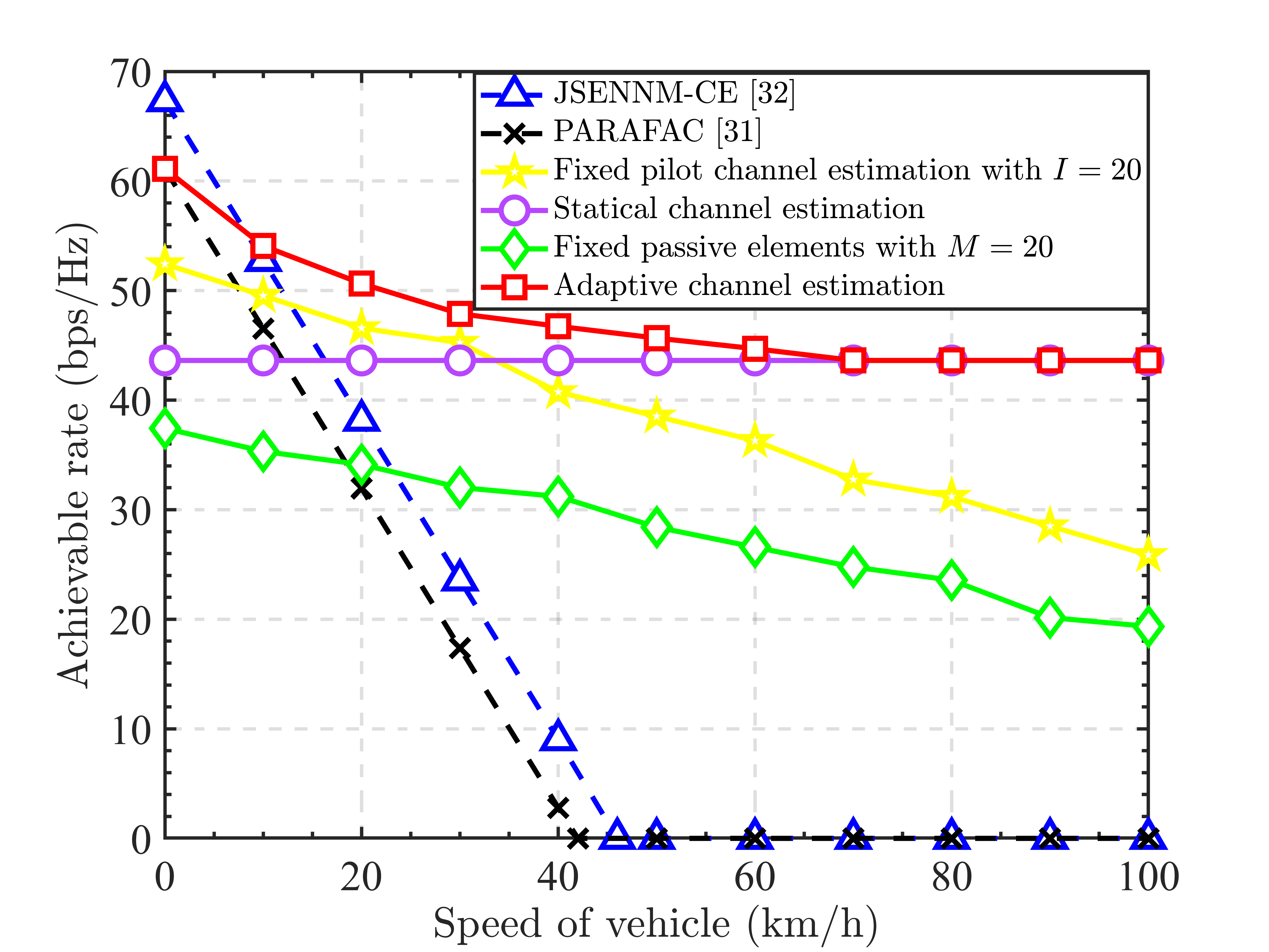}
\vspace{-4mm}
  \caption{Achievable rate versus the speed of vehicle for RIS-aided
 vehicular MIMO communication.}\label{fig6}
\vspace{-6mm}
\end{figure}

In Fig. \ref{fig5}, we show the data rate versus the number of blocks $I$ for the proposed beamforming optimization algorithm  and the benchmark schemes.
It is observed that the performance of algorithms that based on the passive elements grouping method improves with the increase in the number of time blocks.  
Specifically, our proposed algorithm outperforms the statical CSI-based method but underperforms perfect CSI-based method and JSENNM-CE.
It is also observed that as the number of time blocks improves, the performance of the grouped schemes is able to reach the performance of PARAFAC when $I=M$.

Then, to evaluate the robustness of the proposed beamforming algorithm to the VUE travel speed, we define the achievable rate as
$
{R}_a=(1-\frac{T_e}{T_c})R^{\text{nb}}
$,
where $T_e$ and $T_c$ are channel estimation time and channel coherence time, respectively.
With the channel coherence time of $T_c$ and the channel estimation overhead of $T_e$, $1-\frac{T_e}{T_c}$ is the fraction of time that is used for data transmission. 
Thus, the fraction captures the loss in achievable rate due to the channel estimation.
As the vehicle travels faster, the channel changes faster, which leads to a more costly channel estimation overhead.
The root mean square delay spread is set as $\tau_{R\!M\!S}=1 \mu s$.
In Fig. \ref{fig6}, we show the achievable rate versus the VUE speed with the number of blocks and passive elements are fixed at $I=20$ and $M=100$, respectively.
Except in low-speed or quasi-static scenarios, our proposed optimization scheme based on adaptive channel estimation consistently outperforms other benchmark approaches.
When the VUE moves slowly, the algorithm automatically increase the number of pilots to enhance the accuracy of channel estimation, thus improving the transmission performance of the system. 
And when the VUE moves faster, the algorithm will reduce the number of pilots until the performance enhancement brought by improving the channel estimation accuracy is lower than the performance degradation caused by the pilot overhead, and the algorithm will optimize based on statistical CSI directly. 
In summary, the optimization based on adaptive channel estimation has the highest robustness to speed.
It can be observed that except scheme optimized based on statical channel estimation, the achievable rates of all schemes decrease rapidly as the vehicle travels faster.
Due to the low channel estimation error, hybrid beamforming optimization based on JSENNM-CE with $M=100$ shows the best performance in stationary or low-speed scenarios, but the performance degrades fast as the vehicle travels faster, and even collapses at speeds up to about 43 km/h.
Similar results can also be seen on PARAFAC.
In contrast, the scheme with a fixed pilot channel estimation with $I=20$ shows better performance and ensures normal communication even at speeds up to 100 km/h.
To draw more insight, we further compare the fixed pilot $I=20$ scheme with the fixed passive elements $M=20$ scheme. 
In this setup, the channel estimation overhead is the same.
As depicted in the figure, the fixed pilot $I=20$ scheme outperforms the fixed passive elements $M=20$ scheme.
This is because although the number of optimization variables for the optimization objective is the same for both schemes, the former scheme, after group channel-based optimization, the higher number of passive elements can provide additional performance gains for the system.
It is also noted that in low-speed scenarios, the fixed pilot scheme performs better than the statistical CSI-based scheme, while in medium- to high-speed scenarios above approximately 35 km/h, the fixed pilot scheme performs poorly compared to the statistical CSI-based scheme.
This is because the channel estimation overhead advantage of the statistical CSI-based scheme can compensate for the insufficient accuracy of CSI estimation.
Due to the conflict between channel estimation accuracy and pilot overhead, adaptive pilot scheme needs to achieve a trade-off between channel estimation and achievable rate to maximize system performance.
Consequently, the adaptive channel estimation scheme we proposed exhibits stronger robustness to speed. 
At speeds below 70 km/h, the proposed algorithm outperforms the statistical CSI-based scheme.
In high-speed scenarios, the proposed algorithm exhibits performance equivalent to that based on statical CSI. 
The algorithm tends to optimize beamforming using only statistical CSI, as the improvement in channel estimation accuracy cannot compensate for the loss in achievable rate.

Last, we evaluate the performance of the hybrid beamforming algorithm for multi-VUE in broadband OFDM system with VUE number $K=3$ and the distances from VUEs to BS are set as 800 m, 1000 m and 1500 m, respectively.
The bandwidth is set as $B=10$ MHz.
The sub-carrier number is $N=32$.
The QoS threshold is set as $C_{\min}=3\times10^7$ bps.
\begin{figure}{}
\centering
  \includegraphics[width=2.8in,height=2in]{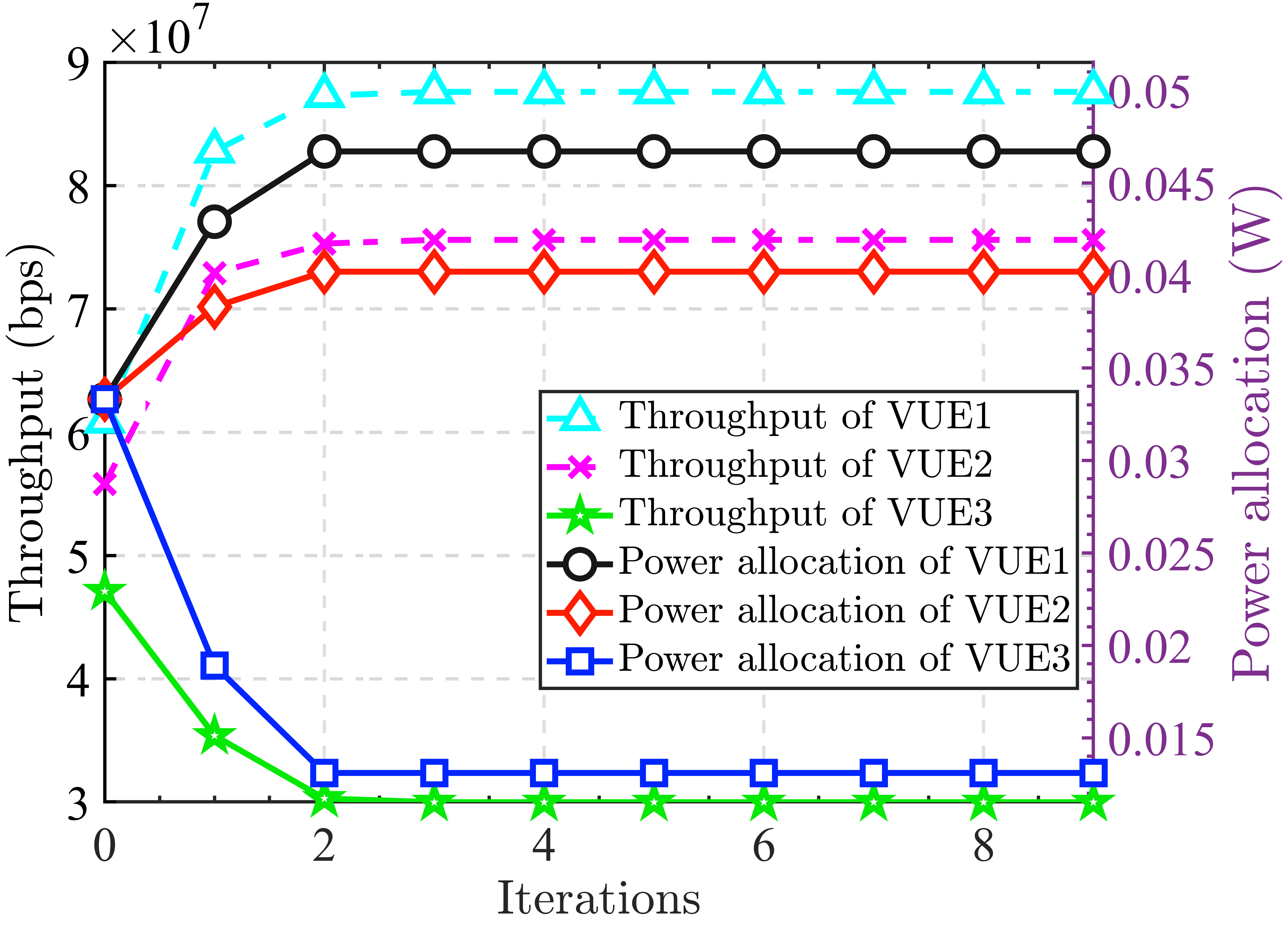}
\vspace{-4mm}
  \caption{Throughput and power allocation of all VUEs versus number of iterations for multi-VUE OFDM system.}\label{fig8}
\vspace{-6mm}
\end{figure}
\begin{figure}{}
\centering
  \includegraphics[width=2.8in,height=2in]{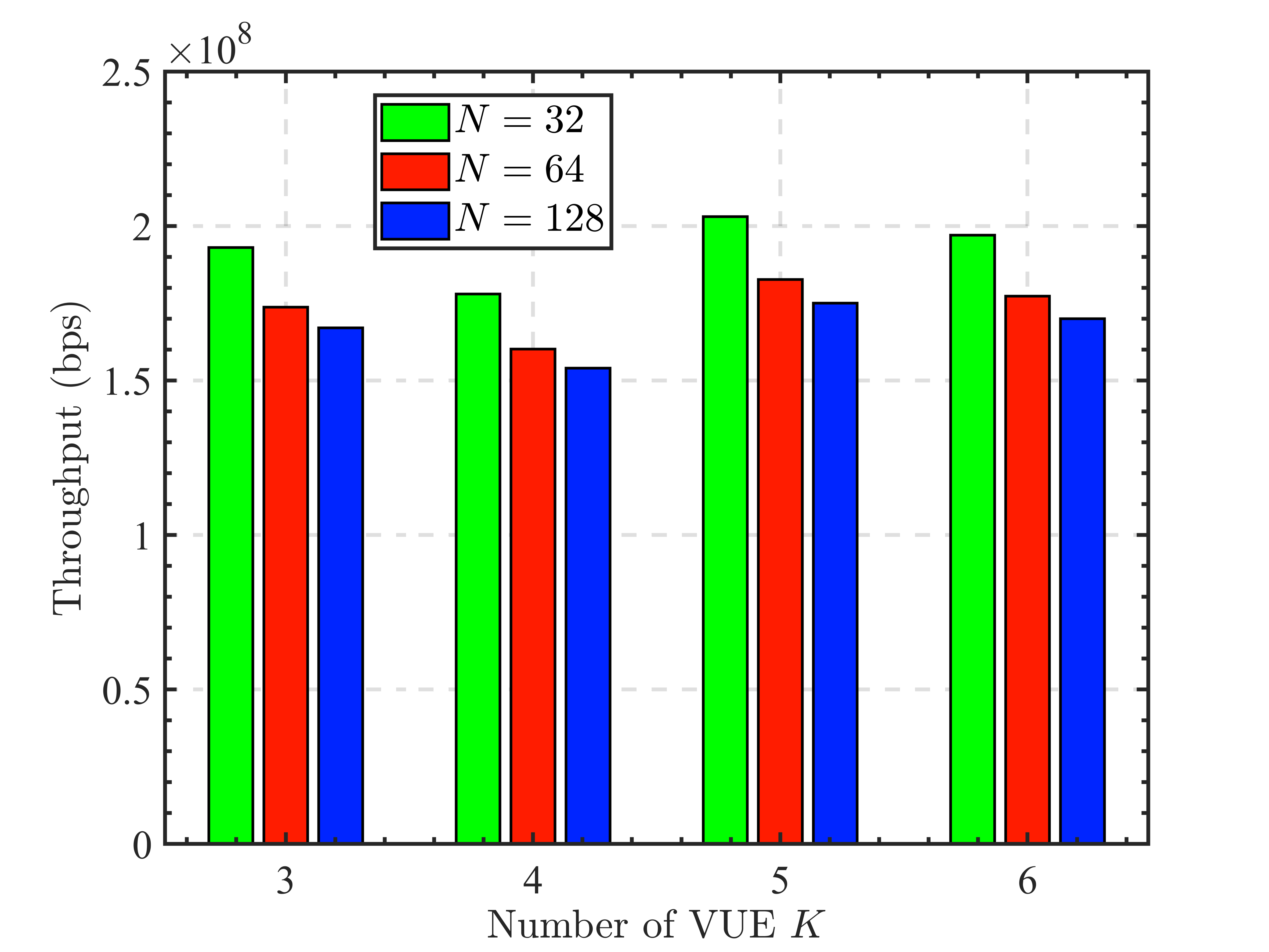}
\vspace{-4mm}
  \caption{Total throughput of different number of sub-carriers versus  VUE number $K$ for multi-VUE OFDM system.}\label{fig9}
\vspace{-6mm}
\end{figure}
We evaluate the convergence behavior of the proposed resource allocation algorithm.
In each iteration, we have to use the proposed algorithm for finding the optimal resource allocation and hybrid beamforming settings of the multi-VUE vehicular communication system.
As depicted in Fig. \ref{fig8}, the throughput of all VUEs converge fast in the first two iterations, and the power allocation does the same.
In particular, VUE3's throughput drops to $3\times 10^7$ due to the minimum QoS restriction and does not decrease further.
This ensures fairness in the resource allocation between VUEs.

In Fig. \ref{fig9}, we investigated the effects of the number of VUEs and the number of sub-carriers on total throughput.
In the proposed system, we are concerned with total throughput and are not concerned with which VUEs are allocated sub-carriers, as long as the channel conditions for VUEs are not significantly poor. 
The more refined allocation resulting from an increase in the number of VUEs will not have a significant impact on total throughput.
And since we ignored the multi-link between multiple RISs and the penetration signals through the vehicle body, the increase in the number of RISs due to the increase in the number of users would not cause additional interference.
Therefore, as the number of VUEs increases, there is no obvious trend in the total throughput.
It is also observed that the system achieves the highest throughput when $N=32$, while throughput decreases progressively as $N$ increases to 64 and 128.
This is mainly attributed to two reasons.
On one hand, the RIS employs a common phase shift configuration across all sub-carriers due to hardware constraints, which lack the ability for frequency-selective passive beamforming. 
As $N$ increases, the diversity in channel characteristics across sub-carriers grows, yet the RIS must still apply a single reflection configuration for each VUE. 
This results in sub-optimal passive beamforming for all sub-carriers, thereby degrading the aggregated signal quality.
On the other hand, for a fixed total system bandwidth, increasing the number of sub-carriers $N$ leads to narrower sub-carrier interval $\Delta f$. 
In vehicular communication system, the ICI becomes more severe due to increased Doppler-induced frequency offsets, which leads to a low SNR.
It is worth noting that when the number of sub-carriers further increases from 64 to 128, although the frequency selectivity continues to increase, the marginal degradation in RIS performance becomes less pronounced. 
This is because the RIS gain has already been substantially diluted over more sub-carriers, and the system approaches a regime where additional sub-carriers yield diminishing returns in terms of RIS-assisted capacity enhancement.
Therefore, the capacity degradation due to increasing sub-carriers from 32 to 64 is more significant than that from 64 to 128. 
\vspace{-4mm}
\section{Conclusion}
\vspace{-1mm}
In this paper, we have proposed an adaptive pilot estimation framework and enhanced the VUE performance for vehicular communication systems by employing a transparent RIS at the top of the vehicle. 
Specifically, by carefully designing the pilot signal, the two time-scale CSI of the cascaded channel can be estimated progressly one block by block.
We firstly studied the hybrid beamforming optimization in narrowband system with single-VUE based on the estimated CSI.
Then we studied the resource allocation and hybrid beamforming optimization in broadband system with multi-VUE, while guaranteeing total power constraint, unit-modulus constraint at the RIS and minimal QoS requirements for all VUEs.
Our simulation results verified that the proposed algorithms performed better than their conventional counterparts.
The adaptive pilot scheme can achieve more robust performance gains in high-speed scenarios.
Moreover, it was revealed that in vehicular OFDM communication system, the number of sub-carriers require to be designed judiciously to achieve the trade-off between ICI, passive beamforming gain and spectrum efficiency.
An interesting future work direction is to investigate the collaboration between multi-VUE and multi-RIS to further improve system performance with practical considerations, such as imperfect/partial CSI knowledge and information security issues in distributed systems.
\appendices
\vspace{-4mm}
\section{Proof of Theorem 1}
According to the property of matrix operation that $(\bm{AB}\otimes\bm{CD})=(\bm{A}\otimes\bm{C})(\bm{B}\otimes\bm{D})$, we can reformulate (\ref{theta_channel}) as
\begin{align}
\bar{\bm{y}}&=\bar{\bm{\theta}}^I\left(\left(\bar{\bm{H}}^I\right)^T\odot\bar{\bm{G}}^I\right)^T\left(\bm{F}\otimes\bm{I}_{N_r}\right)\left(\bm{x}\otimes\bm{I}_{N_r}\right)+\bar{\bm{n}},\notag\\
&=\bar{\bm{\theta}}^I\bar{\bm{H}}^I_e\left(\bm{x}\otimes\bm{I}_{N_r}\right)+\bar{\bm{n}},\notag\\
&=\begin{bmatrix}
\bar{\theta}_1,\bar{\theta}_2,\dots,\bar{\theta}_I
\end{bmatrix}\begin{bmatrix}
\bar{\bm{h}}_e^1\\\bar{\bm{h}}_e^2\\\vdots\\\bar{\bm{h}}_e^I
\end{bmatrix}\begin{bmatrix}
{\bm{x}}_1\bm{I}_{N_r}\\{\bm{x}}_2\bm{I}_{N_r}\\\vdots\\{\bm{x}}_I\bm{I}_{N_r}
\end{bmatrix}+\bar{\bm{n}}.\label{theta_channel2}
\end{align}
Consequently, the formulation can finally be written as
\begin{align}
\bm{y}=\bar{\bm{y}}^T=\left(\sum_{i=1}^{I}\bar{\theta}_i\tilde{\bm{H}}_e^i\right)\bm{x}+\bar{\bm{n}}^T=\bm{\mathcal{H}}\bm{x}+{\bm{n}},
\end{align}
and we can easily get the data rate in (\ref{Capacity2}). This ends the proof.

\vspace{-3mm}
\section{Proof of Lemma 1}
From the formulation given in (\ref{Cnb2}), the optimization problem can be reformulated as the maximization of the following
\begin{align}
g(\alpha)=\det\left(\bm{A}_i+\alpha\bm{B}_i+\alpha\bm{B}^H_i\right),
\end{align}
defined over the unit circle $S=\{\alpha\in\mathbb{C}|\ |\alpha|=1\}$.
To analyze this problem, we firstly express $g(\alpha)$ in terms of a polynomial.
By factoring out $\alpha^{-1}$, we obtain:
\begin{align}
g(\alpha)&=\det\left(\alpha^{-1}\left(\alpha\bm{A}_i+\alpha^2\bm{B}_i+\bm{B}^H_i\right)\right)\notag\\
&=\alpha^{-N_r}\det\left(\alpha^2\bm{B}_i+\alpha\bm{A}_i+\bm{B}^H_i\right).
\end{align}
Let $P(\alpha)\triangleq\det\left(\alpha^2\bm{B}_i+\alpha\bm{A}_i+\bm{B}^H_i\right)$, which is a polynomial in $\alpha$ of degree at most $2N_r$.
The function $g(\alpha)=\alpha^{-N_r}P(\alpha)$ is a Laurent polynomial, 
and its extreme points on $S$ must satisfy the critical condition $Q(\alpha)\triangleq\alpha P'(\alpha) - a P(\alpha) = 0$.
Here, $Q(\alpha)$ is also a polynomial of degree at most $2N_r$. 
Consequently, the optimal $\alpha$ must correspond to a root of $Q(\alpha)=0$ lying on the unit circle $S$.
The solvability of this optimization problem depends critically on the degree $2N_r$. 
For $N_r=1$, $Q(\alpha)$ is quadratic and the closed-form solution exists.
For $N_r=2$, $Q(\alpha)$ is quartic, which is solvable in radicals, though the formulas are complex. 
For $N_r>2$, the degree of $Q(\alpha)$ is $2N_r\geq6$. 
According to the Abel-Ruffini theorem, general polynomial equations of degree five or higher do not admit solutions expressible in radicals in terms of their coefficients. 
Thus, for $N_r\geq3$, there is no general closed-form solution to maximize the objective function (\ref{Cnb2}) using only elementary operations and radicals.
This ends the proof.
\ifCLASSOPTIONcaptionsoff
  \newpage
\fi



%
\footnotesize
\vspace{-6mm}
\bibliographystyle{IEEEtran}
\bibliography{Bibliography}

\end{document}